\documentclass{llncs}
\usepackage{amsfonts}
\usepackage{amssymb}
\usepackage[pdftex]{graphicx}
\graphicspath{{figures/}{../jpeg/}{../png/}}
\DeclareGraphicsExtensions{.jpeg,.png}
\usepackage{mdwmath}
\usepackage[cmex10]{amsmath,mathtools}
\usepackage{algorithmic}
\usepackage{algorithm}
\usepackage[compatibility=false]{caption}
\usepackage{subcaption}
\numberwithin{figure}{section}
\numberwithin{table}{section}
\numberwithin{equation}{section}
\usepackage{subfiles}
\usepackage[sort,nocompress]{cite}
\usepackage{url}
\urldef{\mailsa}\path|{royd,anish}@cse.ohio-state.edu, mukundan.sridharan@samraksh.com, sdeshpande@appdynamics.com|

\DeclareMathOperator*{\argmin}{\arg\!\min}
\begin{document}
  \mainmatter
\title{Achieving Throughput via Fine-Grained Path Planning in Small World DTNs}
\author{Dhrubojyoti Roy\and Mukundan Sridharan\and Satyajeet Deshpande\and Anish Arora}
\institute{Department of Computer Science and Engineering,\\
The Ohio State University,\\
Columbus, OH-43210, USA.\\
\mailsa}

   \maketitle
\documentclass[main.tex]{subfiles}
\begin{abstract}
\label{sec_abstract}
We explore the benefits of using fine-grained statistics in small world DTNs to achieve high throughput without the aid of external infrastructure.  We first design an empirical node-pair inter-contacts model that predicts meetings within a time frame of suitable length, typically of the order of days, with a probability above some threshold, and can be readily computed with low overhead. This temporal knowledge enables effective time-dependent path planning that can be respond to even per-packet deadline variabilities.  We describe one such routing framework, REAPER (for Reliable, Efficient and Predictive Routing), that is fully distributed and self-stabilizing. Its key objective is to provide probabilistic bounds on path length (cost) and delay in a temporally fine-grained way, while exploiting the small world structure to entail only polylogarithmic storage and control overhead. A simulation-based evaluation confirms that REAPER achieves high throughput and energy efficiency across the spectrum of ultra-light to heavy network traffic, and substantially outperforms state-of-the-art single copy protocols as well as sociability-based protocols that rely on essentially coarse-grained metrics.
\end{abstract}
\documentclass[main.tex]{subfiles}
\section{Introduction}
\label{sec_intro}
Disruption Tolerant Networks (DTNs) are characterized by intermittent connectivity and potentially high end-to-end delays. As a consequence, the lack of adequate awareness of temporal path variability impacts throughput, energy efficiency and delay In particular, state-of-the-art DTN routing protocols that use time-averaged metrics and on-the-fly forwarding decisions based on past history not only miss out on good forwarding opportunities but also are vulnerable to using paths of poor quality. The more recent sociability-based protocols expend large control overheads to achieve better path planning, but are insensitive to variable packet deadlines. In contrast, we explore in this paper whether a DTN protocol with an expectation of future contacts in a ``prediction time frame''---a window into the future---can effectively employ time-dependent path-planning.  We leverage recent analyses of several mobility traces\cite{PowerLaw07, Dichotomy07, Pairwise07} that have shown periodic behavior being exhibited along 
with some degree of randomness.  These hint at the possibility of achieving fine-grained predictability of node inter-contacts within some time frame and, thereby, making time-dependent choices of paths which are reliable and/or timely.  Recent research has also demonstrated a ``small-world phenomenon'' in real-life dynamic networks\cite{Watts98, Kleinberg00, Nguyen12}, wherein a network is connected via short paths that grow only logarithmically with its size.  These suggest that as long as frame based, fine-grain path planning preserves the small-world property, enumeration of available paths should not incur high overhead.

{\bf Contributions of the paper.}~~~We introduce a temporal frame-based approach for DTN routing.  Intuitively, a frame is an interval of time in which node-pair inter-contact (link) probabilities are high for sufficiently many pairs such that together these pairs imply a temporally giant connected graph, over which reliable routing is possible even if there is a trade-off with delay.  Through the analysis of real life dynamic network traces, we show that a suitably chosen time frame allows for deterministic or highly reliable meetings of a significant fraction of node pairs in the network, greatly improving the average path probabilities and expected throughput, while preserving the small-woorld property of the network. In the context of a frame, we next introduce a model of pairwise inter-contacts that can be used for predicting upper-bounds on node meeting instants. The model is simple to implement, and requires minimal information in a real-world dynamic setting. It does not assume the existence of 
infrastructure nor any advance knowledge of the underlying mobility model and network dynamics: nodes leverage the intrinsic pattern of inter-contacts to extract link information in the course of normal network evolution.

Using the predicted schedules, we then demonstrate fine-grained network path planning using traditional QoS metrics (such as path length, delay, reliability, etc.), by designing REAPER (for Reliable, Efficient and Predictive Routing), a fully distributed single-copy routing framework that allows for temporally variable selection of the best path subject to some deadline constraints for message delivery. The common case control overhead of REAPER is polylogarithmic in the number of relay nodes, while maintaining enough information to be used with a variety of optimization objectives (in this paper, we optimize for path lengths, incerchangeably referred to as costs). REAPER is flexible in that it allows for the specification of the delay constraint on a per-message basis. The protocol also self-stabilizes under a bounded number of failures in the system; the formal proof of stabilization is presented in the Appendix.

We have evaluated the performance of REAPER against link-metric based infrastructure-free single-copy DTN protocols, namely, Minimum Estimated Expected Delay-Based Distance Vector Routing (MEED-DVR) \cite{MeedDvr07} and Probabilistic Routing Protocol using a History of Encounters and Transitivity (PROPHET) \cite{Prophet04}. The evaluation is based on ns-2 simulations and considers a realistic human-carried mobile network setting. It shows an improvement in throughput of up to $135\%$ ($200\%$) over MEED-DVR (PROPHET) across light to heavy traffic loads, with up to $4.5\times$ ($22\times$) lesser average path lengths.  In addition to discussing how these improvements hold even with respect to more recent single-copy protocols, we observe that REAPER has $\mathcal{O}(\log{}n)$ (best case $\mathcal{O}(n)$) lower control overhead than the sociability metric based BUBBLE Rap\cite{Bubble11}, and has about $10\%-40\%$ better throughput as a consequence of fine-grained path planning.

{\bf Organization of the paper.}~~~In Section \ref{sec_relwork}, we present related work on small worlds, inter-contact modeling, and routing in DTNs.  In Section \ref{sec_sysmodel}, we introduce and analyze the frame-based inter-contacts model and prediction problem.  We describe REAPER and provide simulation-based validation in Sections \ref{sec_routing} and \ref{sec_expt} respectively. Section \ref{sec_conclusion} features concluding remarks and discusses future work. Finally, the Appendix section presents the complete protocol and provides proofs of self-stabilization, correctness, optimality, and cycle-avoidance.
\documentclass[main.tex]{subfiles}
\section{Related work}
\label{sec_relwork}

{\bf Small worlds in dynamic networks.}~~~As early as 1967, Milgram demonstrated in a famous experiment that human acquaintances are connected by the proverbial ``six degrees of separation''. Until relatively recently, research on the small world phenomenon predominantly analyzed static graphs \cite{Watts98,Kleinberg00}. However, recent research has shown this phenomenon manifests in dynamic graphs as well. Through the existence of spatio-temporal ``communities'' ``superconnectors'' that carry information across communities, the small-world phenomenon results in networks with only polylogarithmic path lengths. Nguyen et al \cite{Nguyen12} demonstrated high correlations of shortest dynamic path lengths and clustering coefficient over periods of 24 hours, attributing the observations to cycles in daily human activities. Clauset et al \cite{Clauset07} and Nguyen et al \cite{Nguyen13} designed autocorrelation-based metrics for measuring temporal similarity and estimating the extent of disorder in small world 
dynamic 
networks. It turns out however, for reasons we discuss in Section \ref{sec_sysmodel}, that autocorrelation techniques yield poor estimators for characterizing pairwise inter-contacts.

{\bf Routing in DTNs.}~~~Routing algorithms in DTN literature can be classified into two broad categories: replication-based and single-copy. In the former category, some prominent examples are Epidemic \cite{Epidemic00}, Spray and Wait \cite{SprayWait05}, MaxProp \cite{MaxProp06} and RAPID \cite{Rapid07}. Backpressure-based routing schemes use queue differentials to make forwarding decisions that achieve throughput optimality. A recent variation of backpressure routing, BWAR \cite{BWAR12}, jointly optimizes throughput and delay, but it relies on the ability to forward commodities destined for some node at every link and therefore reduces to epidemic-based flooding for a single destination class.

Our work belongs to the second category of protocols, which uses statistical link metrics to choose a suitable relay for single-hop forwarding. This approach inherently yields more energy efficiency, if not more goodput, but its throughput or delivery delay can be adversely affected if the forwarding node is not chosen properly. Popular infrastructure-free benchmark protocols in this category are PROPHET \cite{Prophet04} and MEED \cite{MeedDvr07}, which respectively maximize path probability and minimize expected delay, and serve as the basis for comparison with REAPER.  Other probabilistic forwarding algorithms, such as \cite{Per09, OptimalProbForwarding09} assume some kind of global knowledge or the existence of static infrastructure. In the context of social networks, a third category of protocols has found prominence in recent times: those that route packets based on sociability information. BUBBLE Rap\cite{Bubble11} is such a benchmark that performs quite well in terms of throughput and cost in small-
world networks, and is known to perform better than other centrality-based protocols (e.g., Simbet\cite{Simbet}). In this paper, we analyze  how fine-grained path planning gives our framework higher throughput and much better control overhead than BUBBLE Rap. More recent work such as\cite{Gao10,Zhang13} advocate the discovery of temporally transient contact patterns and communities from DTN traces; however, such methods require global knowledge and are realized in a purely centralized manner.
\documentclass[main.tex]{subfiles}
\section{Frame-Based Reliability and The Prediction Problem}
\label{sec_sysmodel}
\subsection{Finding characteristic time frames in Human-Centric DTNs}
\begin{figure}[t]
\centering
            \begin{subfigure}[b]{0.35\textwidth}
                    \includegraphics[width=\textwidth]{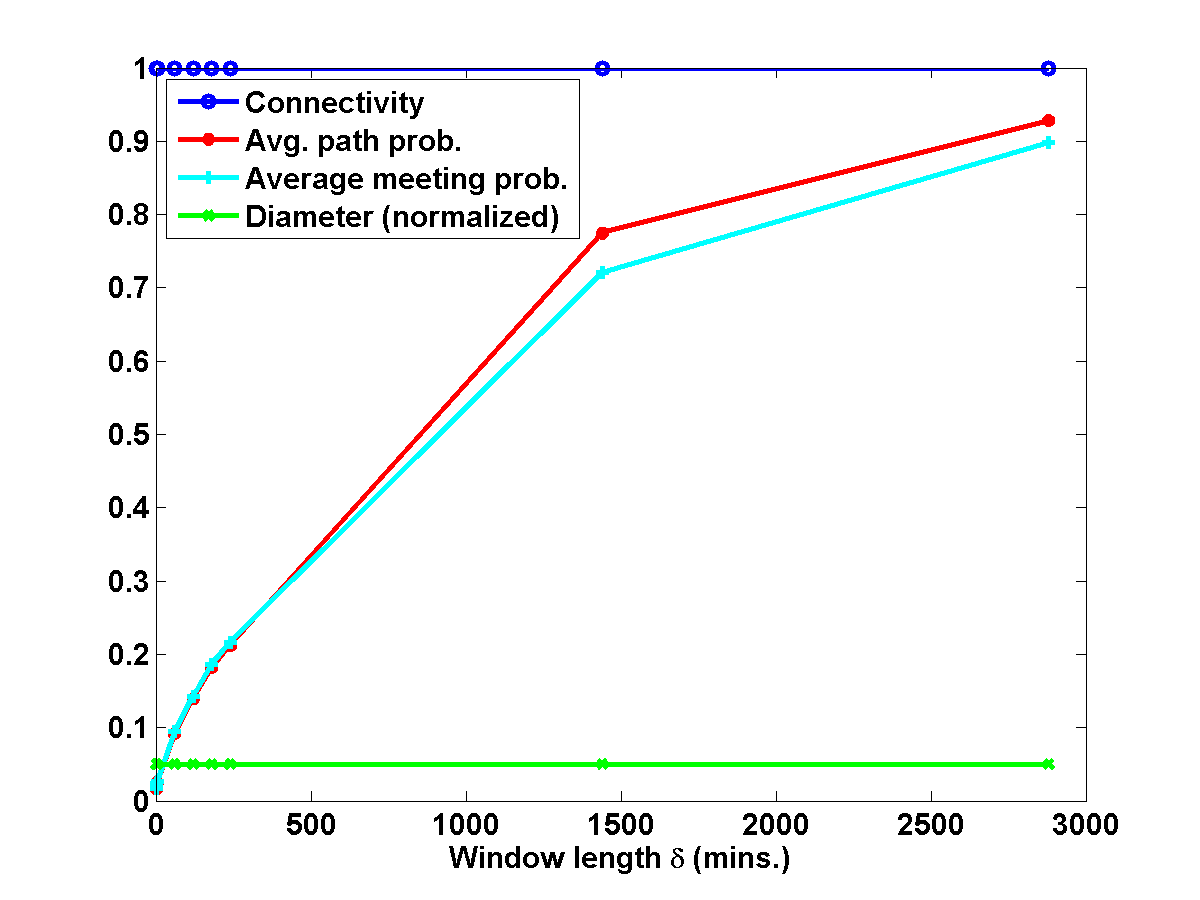}
                    \caption{Infocom05}
            \end{subfigure}
            \begin{subfigure}[b]{0.35\textwidth}
                    \includegraphics[width=\textwidth]{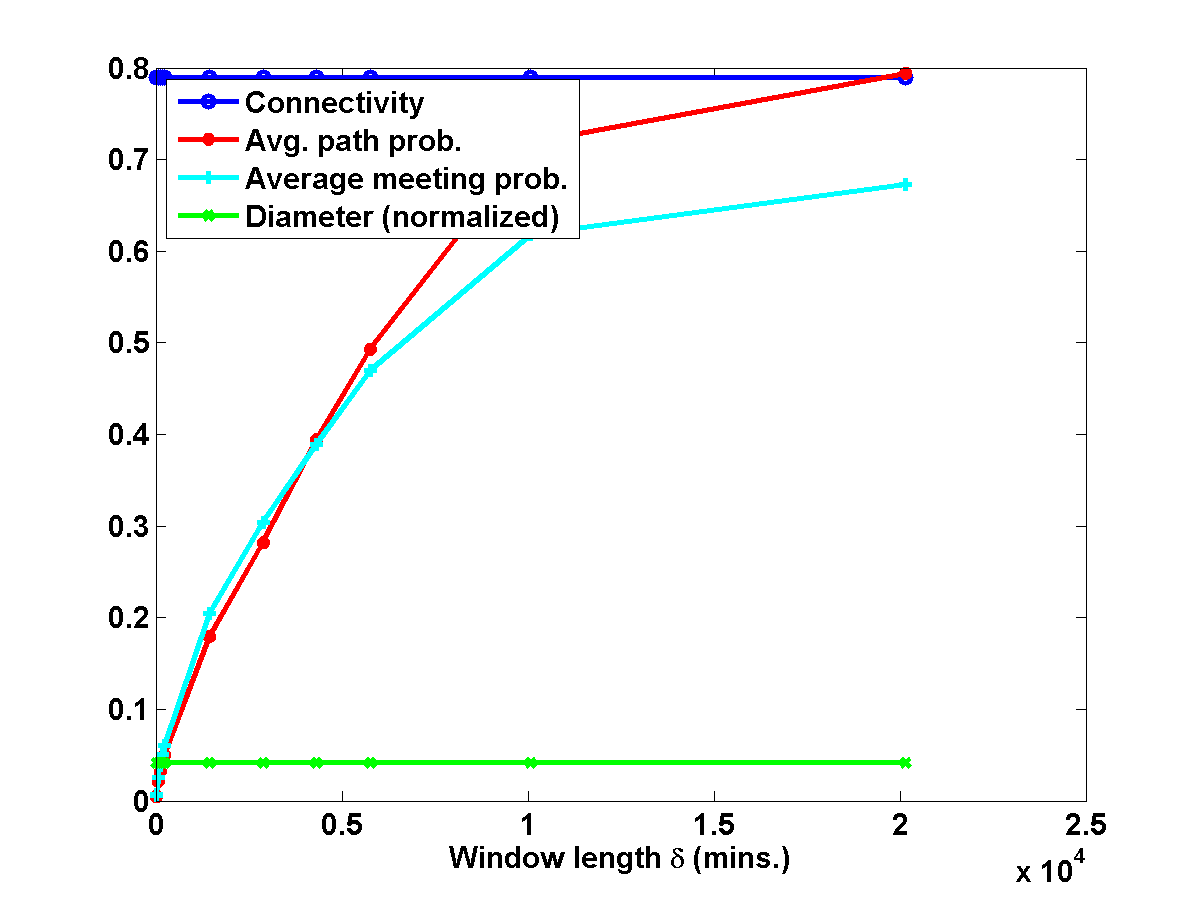}
                    \caption{Milan}
            \end{subfigure}

            \begin{subfigure}[b]{0.35\textwidth}
                    \includegraphics[width=\textwidth]{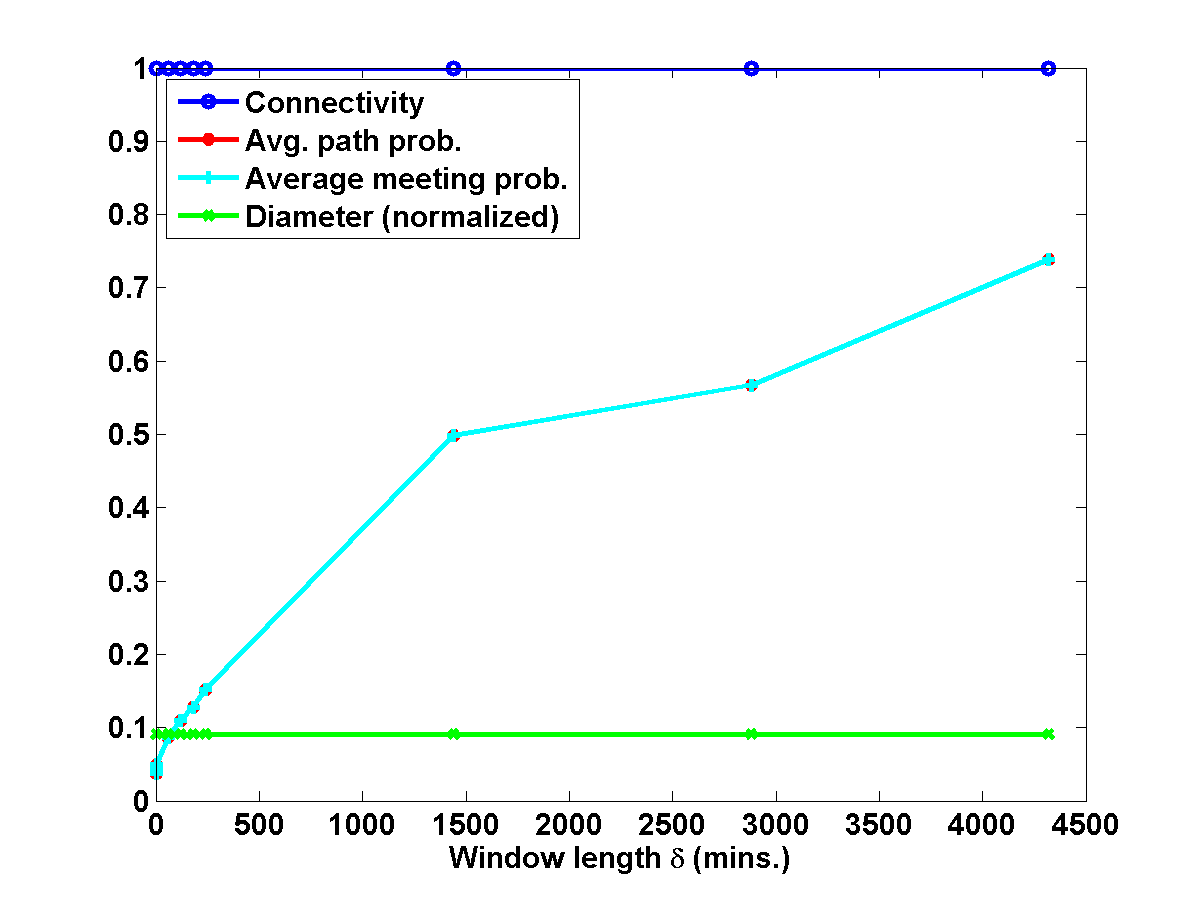}
                    \caption{Cambridge05}
            \end{subfigure}
            \begin{subfigure}[b]{0.35\textwidth}
                    \includegraphics[width=\textwidth]{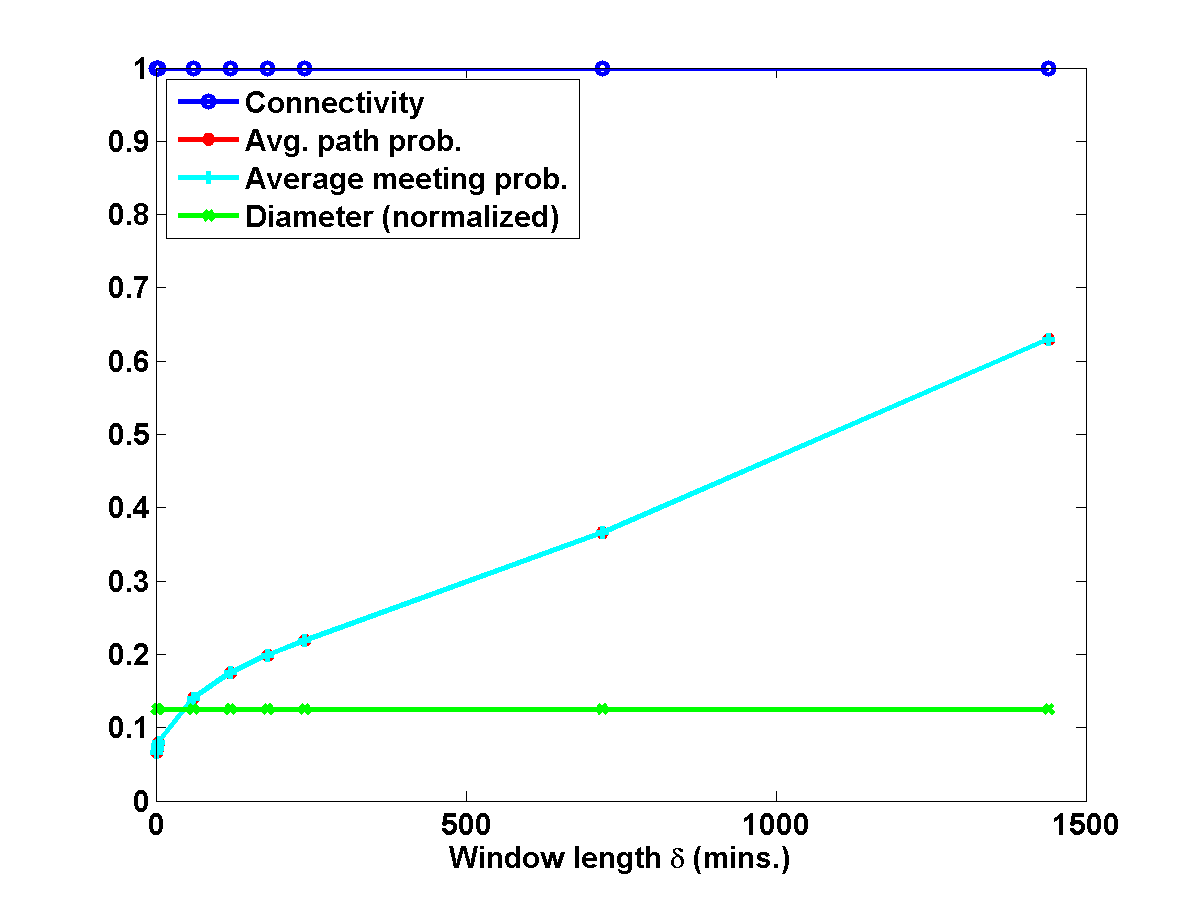}
                    \caption{Intel}
            \end{subfigure}
            \caption{Normalized metrics with varying window size $\delta$}\label{fig_windows}
\end{figure}
We use three data sets from the Dartmouth Haggle project\cite{haggle06} and one from the University of Milano\cite{pmtr08} that contain opportunistic Bluetooth contact traces between users in conference, lab or campus environments. In order to focus on purely peer-to-peer communication, we filter out all contacts with access points as well as with external nodes that can facilitate message transmission by acting as static or mobile infrastructure. The traces are analyzed as follows: we iterate over the interval duration, $\delta$, from $1$ minute to roughly half of the total trace duration $T$. In each iteration, we consider contiguous windows of length $\delta$ and calculate the following metrics of interest: (i) the average per-node inter-contacts probability, (ii) the percentage of connected links, (iii) the average path probability considering the most reliable paths, and (iv) the network diameter comprising of shortest hop paths.

The normalized metrics are presented in Fig. \ref{fig_windows}(a)-(d). Observe that the node inter-node meeting probability increases greatly with the use of larger time-frames without compromising network connectivity. Interestingly, in each case, the average path quality is at least that of the average link quality in a suitably chosen window of time, despite the fact that path probabilities are multiplicative. This can be inferred from the small world property of the networks in the following way. A fraction of links are assured $100\%$ connectivity within each frame. Some of these relays are also superconnectors, i.e., they communicate with many other nodes in the network, even if with reliability lower than $100\%$, yielding higher path qualities that surpass the individual link probabilities between the corresponding source-destination pairs.
\begin{table}[t]
\caption{Summary of Mobility Traces} 
\centering 
\begin{tabular}{c c c c c c c} 
\hline\hline 
Dataset & Nodes & Environ. & Frame & Avg. & Connect. & Diameter\\  
 & /Days & & (days) & Quality & & (hops)\\[0.5ex]
\hline 
Infocom05 & 41/4 & Conference & 1 & 0.77 & 1 & 2\\ 
Milan & 49/19 & Campus & 6 & 0.72 & 0.98 & 3\\
Cambridge05 & 12/6 & Lab & 3 & 0.74 & 1 & 1\\
Intel & 9/6 & Lab & 1 & 0.63 & 1 & 2\\
[1ex] 
\hline 
\end{tabular}
\label{data_summary} 
\end{table}
It follows from this analysis that the use of time frames for routing in DTNs greatly enhances path quality and, as a result, throughput. The choice of a characteristic frame duration for a certain network is somewhat subjective, and as such trades reliability for overhead. For our purpose, we choose the minimum frame length that yields an average path reliability of $60\%$ or higher while maintaining a giant-connected component of $90\%$ or more of all \emph{available} paths\footnote{Observe that per pair of nodes we could have used a potentially different frame length for the same probability threshold due to variances in their meeting patterns over time. To avoid analysis over common multiples of frame lengths associated with node pairs in paths, we select a ``one size fits all'' characteristic frame length.}. The characteristic frame lengths estimated for the four traces, along with their average path quality, connected component size (\%) and network diameter are shown in Table \ref{data_summary}.
\begin{figure}[t]
\centering
            \begin{subfigure}[b]{0.35\textwidth}
                    \includegraphics[width=\textwidth]{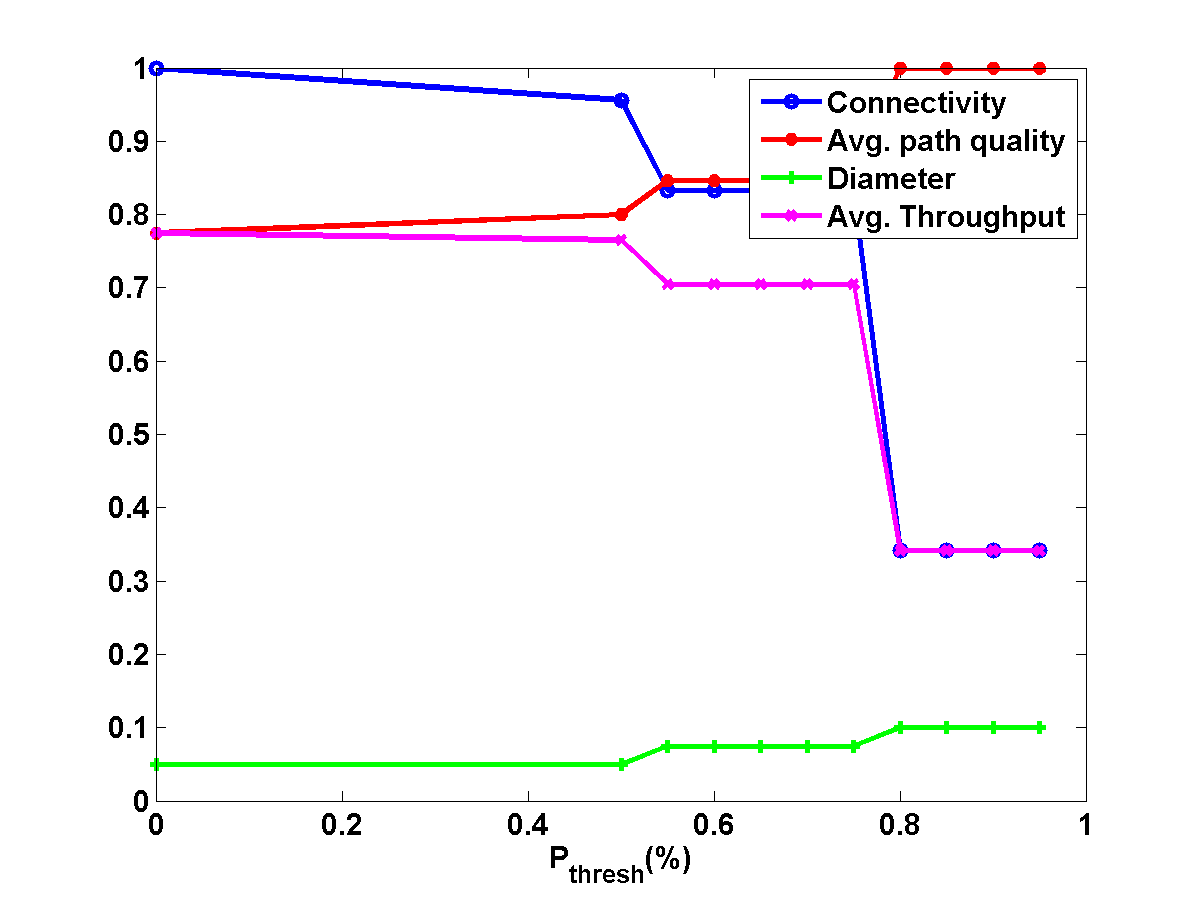}
                    \caption{Normalized metrics}
            \end{subfigure}
            \begin{subfigure}[b]{0.35\textwidth}
                    \includegraphics[width=\textwidth]{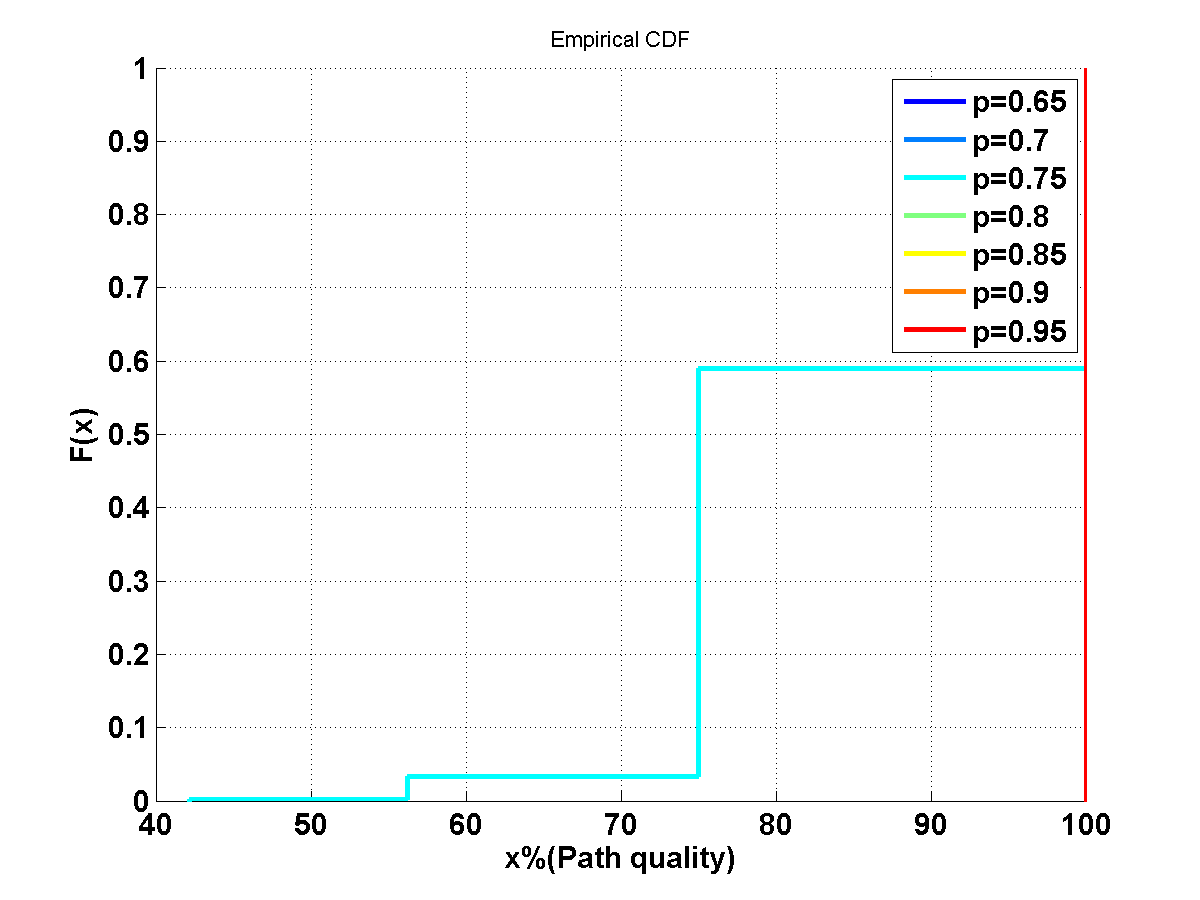}
                    \caption{Empirical CDF}
            \end{subfigure}
\caption{Normalized metrics and path quality ECDF with varying $P_{thresh}$ in Infocom05 ($\delta=1$ day)}
\label{fig_infocom}
\end{figure}
Note that a larger frame length does not preclude the existence or discovery of short-delay paths in the network; it is merely used to enumerate future paths at any given point in the frame to enable routing at a finer level of granularity, as will be discussed in the next section.

The criterion for selecting a frame length raises an important question: for any given frame length, isn't it preferable to only include links of high reliability (say, above a certain cost threshold, $P_{thresh}$) to create paths in the network?  Contrary to intuition however, our analysis shows otherwise. By way of example, Fig.~\ref{fig_infocom}(a) summarizes the variations of network metrics with increasing values of $P_{thresh}$ for the Infocom05 trace using its characteristic frame duration of $1$ day.  Observe that the path reliability increases with increasing $P_{thresh}$, and, while the diameter increases, it still preserves a small world. However, connectivity worsens, and the overall network throughput obtained from Little's Law decreases. This phenomenon may be explained as follows: the $35\%$ assured contacts observed at the threshold of $0.95$ continue to provide the best available paths even when relatively low reliability links are admitted. Thus, even though a direct link between two nodes 
$i$ and $j$ may be admitted with a lower threshold, the best quality path for routing packets between them would likely still be the existing slightly longer path through the assured set of links, since most of the network is already connected through the small-world by then. Consequently, it can be observed in Fig. \ref{fig_infocom}(b) that the path quality ECDFs are identical for a wide range of lower admission thresholds. Lowering the threshold further would help improve connectivity but not affect the existing best paths. When lower quality paths are omitted, the gain in average path quality is surpassed by the loss in connectivity, and consequently, the highest throughput is observed at $P_{thresh}=0.0$. Therefore, the inclusion of low-quality links in a small world network typically enhances its performance once the characteristic frame length has been fixed. A vital difference in our analysis methodology from previous work is worth addressing here. In Nguyen et al \cite{Nguyen12}, an adjacency matrix-based technique for discovering higher-hop path is described. Let $A_t, t=0,1,\hdots,n$ be the adjacency matrix of a dynamic network up to time $t$. According to Nguyen et al, the union $A_t\lor A^2_t\lor\hdots\lor A^n_t$ should enumerate the higher hop paths of length $1,2,\hdots,n$. We regard this estimation as being too optimistic for frame-based computation of higher-hop paths, since the relative ordering of links in time is not captured in the adjacency matrix corresponding to a frame interval. We therefore eschew the occurrence of multiple hops during a frame interval $\delta$, and discover higher-hop paths by computing $A_\delta\lor(A_{(0,\delta]}\cdot A_{(\delta,2\delta]})\lor\hdots\lor((\prod\limits_{x=\delta}^{\lceil \frac{T}{\delta}\rceil-1} A_{(x-1,x]})\cdot A_{(\lceil \frac{T}{\delta}\rceil-1,\lceil \frac{T}{\delta}\rceil]})$. The network diameter, as 
expected, is bounded by as many intervals as the number of logical $\lor$ operations needed for convergence in the number of paths.

\subsection{The Frame Based Inter-Contacts Model}
As explained previously, the node-pair inter-contacts relative to a time frame are essentially a mix of assured and probabilistic events. We propose the following inter-contact model to characterize both aspects for a pair of nodes $i$ and $j$ relative to a frame of length $f$:\\
\noindent
i) \emph{Non-zero, independent meetings.} There exists $z\geq 1$ for each pair of nodes $i$ and $j$ such that the number of meetings in each frame is $z$. The meetings are assumed to be independent.\\
\noindent
ii) \emph{Bounded jitter.} There exists a monotonically non-decreasing sequence of time instants in each frame, $\langle\beta_0,\beta_1,\beta_2,\hdots,\beta_z\rangle$ with $\beta_0=0$, the start of the frame, such that for the first $z-1$ meetings, the $l$\textsuperscript{th} meeting instant between $i$ and $j$ has p.d.f. defined as $p_l(t)$ in $(\beta_{l-1},\beta_l]$. $\beta_z$, which represents the last meeting, however, is allowed to be probabilistic. Mathematically,
$$[\forall l\in[1,z):\int\limits_{\beta_{l-1}}^{\beta_l}p_l(t)dt=1],[\int\limits_{\beta_{z-1}}^{\beta_z}p_l(t)dt\leq 1],[\int\limits_{\beta_{z}}^{f}p_l(t)dt=0]$$
Note that the value of $z$ can be $1$, accommodating the possibility of links with purely probabilistic contacts. The proposed inter-contacts model can be easily extended to slotted time frames, with summations replacing integration. Henceforth in the paper, we divide time into equidistant slots of length $s$ each and use $f$ contiguous slots to constitute the frame between a pair of nodes $i$ and $j$.

\subsection{The Prediction Problem}
Let $H_{ij}$ be a $h\times f$ data structure that enumerates the history of independent contacts between $i$ and $j$ in the past $h$ time frames. The indicator variable $H_{ij}(k,r)$, for $k\in [1,h]$ and $r\in [1,f]$, is $1$ if and only if $i$ and $j$ met at the $r$\textsuperscript{th} slot of the $k$\textsuperscript{th} timeframe.

\noindent 
{\bf The prediction problem can now be stated as:} Given $H_{ij}$ and the present time instant $x$, predict the maximum delay $d_m(x)$ before the next contact between $i$ and $j$.

To solve the prediction problem in our inter-contact model, we first estimate the meeting interval upper-bounds $\beta_1, \beta_2,\hdots, \beta_z$ in the timeframe from history data, essentially converting the set of history timeframes into a single summary data structure that we term as the \emph{$\beta-$frame}. Once the $\beta-$frame has been constructed, the time instant $x$ is mapped to an appropriate position in the $\beta-$frame and the nearest $\beta-$slot is returned.

\noindent
{\bf Step 1: Estimation of slot probabilities.} We first find $\overline{c}_{ij}(r)$, the average number of a contacts per slot for the $r$\textsuperscript{th} timeslot in the timeframe.

\noindent
{\bf Step 2: Estimation of $\beta_1,\beta_2,\hdots\beta_z$.} The $\beta-$slots are estimated by a two-pass method. In the first pass, we keep adding consecutive slot probabilities $\overline{c}_{ij}(r)$ till we reach a slot where the aggregate crosses $1$. If such a slot is found, we mark it as $\beta_1$, and continue to scan for successive $\beta-$slots in the same way, adding any extra probability that may be carried forward from the previous $\beta-$slot. That is, each subsequent $\beta_l$ for $l=2,\hdots,z-1$ is essentially
$$\beta_l=\argmin_t\{(\sum\limits_{r=\beta_{l-2}}^{\beta_{l-1}}\overline{c}_{ij}(r)-1)+\sum\limits_{r=\beta_{l-1}}^{t}\overline{c}_{ij}(r)\geq 1)\}$$
Finally, we represent $\beta_z$ as the last slot observed in $\overline{c}_{ij}$ with non-zero probability. 

The $\beta-$frame constructed in this way is essentially a conservative (upper-bounded) representation of the meeting pattern between a pair of nodes, to be used as input to the routing framework described in the next section. We henceforth use the terms ``frame'' and ``$\beta-$frame'' interchangeably. The present instant $x$ can be normalized to the frame by the formula: 
\begin{equation}
\label{eqn1}
x^\prime=\lceil((x \mod (f\cdot s))/s)\rceil
\end{equation}
The maximum delay before the next contact is, therefore,
\begin{equation}
\label{eqn2}
d_m(x) = \left\{
  \begin{array}{l l}
    (\beta_r-x^\prime)\cdot s+(s-(x\mod s))\text{,} & \\
    \hfill{\text{if } (\exists r: \beta_{r-1}\leq x^\prime < \beta_r),}\\
    ((f-x^\prime)+\beta_r)\cdot s+(s-(x\mod s))\text{,} & \\
    \hfill{\text{otherwise.}}
  \end{array} \right.
\end{equation}

Note that the prediction strategy can compute a fine-grained meeting schedule without explicitly estimating the statistical characteristics of the inter-node mobility model. It can thus be readily computed online and can adapt to changes in meeting patterns given the same frame length. Previous work such as Nguyen et al \cite{Nguyen13} has advocated the use of autocorrelations to find the characteristic frame in a network, by computing autocorrelations over the inter-contact time series for each pair of nodes represented as a bit vector, and selecting the modal interval. The autocorrelation method, however, is restrictive in the sense that it looks for exact matches of corresponding elements in two consecutive intervals. Therefore, it does not capture meetings within bounded jitters, which is the property desired for path planning with bounded QoS estimates.

\documentclass[main.tex]{subfiles}
\section{Reliable Efficient and Predictive Routing (REAPER)}
\label{sec_routing}
In this section, we show how $\beta-$frames can be used to take temporally fine-grained routing decisions with an efficient routing framework that allows for diverse optimization objectives. For ease of exposition, we assume the network consists of $N$ nodes that each seek to send packets to one class of destination nodes (each node in this class is indistinguishably labeled $D$, the ``base station''). That said, the framework is readily extended to support multiple destination classes, anycast, or any-to-any routing requirements. Informally, our specific routing objective is to design a protocol that optimizes the cost of routing to a base station in a disruption tolerant network subject to deadline constraints.  We use the standard notion \cite{Nguyen12} of dynamic path, delay, and cost or path length (interchangeably referred to as ``hop count''). As the destinations are implicit, we abbreviate a path from $i$ to a base station at time $t$ as $L_i(t)$. Also, for ease of exposition, we define two additional functions:\\
\noindent
(i)  \emph{Next-hop, $\xi(L_i(t),j)$}, for node $j$ in path $L_i(t)$ : the node following node $j$ in the path $L_i(t)$.\\
(ii) \emph{Next-contact, $\psi(L_i(t),j)$}, for node $j$ in path $L_i(t)$: the contact time between $j$ and $\xi(L_i(t),j)$ on path $L_i(t)$.

Let $M_{ij}$ be the $\beta-$frame for any pair of nodes $i$ and $j$ of length $F$. {\bf The routing problem can be stated as:} Given a time slot $x$ and a deadline $\Delta$, letting $\Lambda_i(x)$ denote the set of paths from $i$ to D at time $x$, select a path $\widehat{L}_i$ from $\Lambda_i(x)$ that satisfies the optimization objective:
\vspace*{-0.5mm}
\begin{equation}
\label{eqopt}
\begin{split}
 & \theta(\widehat{L}_i)= \min\{\lambda:\lambda\in\Lambda_i(x):\theta(\lambda)\}\text{, subject to}\\
&(\delta(\widehat{L}_i)\leq\Delta) \land (\forall\lambda\in\Lambda_i(x):\theta(\lambda)=\theta(\widehat{L}_i):\delta(\widehat{L}_i)\leq\delta(\lambda))
\end{split}
\end{equation}

The REAPER routing framework uses a fully distributed dynamic programming algorithm for solving the constrained optimization problem via the exchange of control packets between nodes. Each node in REAPER maintains a routing table of $K=\mathcal{O}(log{}N)$ rows called the \emph{t-frame}, as defined next.\\

\textit{\textbf{Definition.}} A \emph{t-frame} $T_i$ at node $i$ is an $K\times F$ matrix of tuples. When routing converges, for $0 < q \leq K$ and $0 < p \leq F$, $T_i(q,p)$ is the tuple $\langle \delta(\widetilde{\lambda}_i),\xi(\widetilde{\lambda}_i,i)\rangle$, where $\widetilde{\lambda}_i$ is a path in $\Lambda_i(p)$ with $\theta(\widetilde{\lambda}_i)=q$ such that
$$\delta(\widetilde{\lambda}_i)=\min\{\lambda\in\Lambda_i(p):(\psi(\lambda,i)=p)\land(\theta(\lambda)=q):\delta(\lambda)\},$$
\noindent
and each $T_i(q,p)$ additionally satisfies the order enforcement property that the stored path delays are always in decreasing order of path lengths:
$$(\forall q^\prime:q^\prime>q:T_i(q^\prime,p)\neq\varnothing\implies T_i(q^\prime,p).\delta<T_i(q,p).\delta),$$
\noindent
where $T_i(q,p).\delta$ and $T_i(q,p).\xi$ respectively refer to the first and second components of $T_i(q,p)$.

\begin{algorithm}[t]
\caption{Constructing $S_{ij}$: best paths for $j$ through $i$}
\label{alg2}
\begin{algorithmic}
\FOR {$q=2\hdots K$}
\FOR {$m=1\hdots z$}
\IF {$\beta_m<\beta_z$}
\STATE $S_{ij}(q,\beta_m):=\min\{r:r\in(\beta_m,\beta_{m+1}]\land(T_i(q-1,r).\xi\neq j):T_i(q-1,r).\delta+(r-\beta_m)\};$
\ELSIF {$\beta_m=\beta_z$}
\STATE $S_{ij}(q,\beta_m):=\min\{$\\
\STATE \qquad $\min\{r:r\in(\beta_m,F]\land(T_i(q-1,r).\xi\neq j):T_i(q-1,r).\delta+(r-\beta_m)\},$\\
\STATE \qquad $\min\{r:r\in[1,\beta_1]\land(T_i(q-1,r).\xi\neq j):\{r+(F-\beta_m)+T_i(q-1,r).\delta\};$
\STATE $\}$
\ENDIF
\ENDFOR
\ENDFOR
\STATE Enforce descending order of delays down a column
\end{algorithmic}
\end{algorithm}

Thus, a t-frame has one or more entries corresponding to each of the $F$ slots, where the entry $(q,p)$ stores the best-delay $q$-cost path and the corresponding next-hop neighbor originating at slot $p$ (which must be a meeting slot corresponding to the said neighbor). Note that the same $p$-slot can offer a separate best path through the same or different neighbor(s) corresponding to each possible path length from $1$ to $K$. However, we only store such paths if a higher-cost path has strictly lesser delay than all lower-cost paths in the same slot.

A node $i$ that meets a destination node enlists $\langle 0,D\rangle$ corresponding to each meeting slot in the t-frame. It then propagates to each meeting node $j$, via a data structure called the s-frame, the path information from its t-frame adjusted relative to its meeting schedule with $j$. To construct the s-frame, $i$ first ``aligns'' the slots of its t-frame with those in the $\beta-$frame $M_{ij}$ (a one-to-one correspondence is observed since both frames are of the same length $F$). $i$ then computes, corresponding to each meeting slot $\beta_m, \forall m\in[1,z]$, the best delay path available through it at each hop between the t-frame slots in the range $[\beta_m+1,\beta_{m+1}]$ that does not have $j$ as the next neighbor (for best-effort cycle avoidance).\\

\textit{\textbf{Definition.}} An \emph{s-frame} $S_{ij}$ is an $K\times z$ matrix that advertises from $i$ to $j$ the best-delay paths $j$ has via $i$ at the meeting times in $M_{ij}$. When the routing algorithm converges, with $0 < q \leq K$ and the column entry ranging over the slots $\beta_1\hdots\beta_z$ in $M_{ij}$, each element $S_{ij}(q,\beta_m)$ is the singleton $\langle\delta(\widetilde{\lambda}_j)\rangle$, where $\widetilde{\lambda}_j$ is a path in $\Lambda_j(\beta_m)$ with $\theta(\widetilde{\lambda}_j)=q$ such that
\begin{equation}\nonumber
\begin{array}{l l}
\delta(\widetilde{\lambda}_j)=\min\{\lambda\in\Lambda_j(\beta_m):(\xi(\lambda,j)=i)\land(\xi(\lambda,i)\neq j)\land
&\\
\hfill{(\theta(\lambda)=q-1)\land(\psi(\lambda,i)\in R):\delta(\lambda)\}}
\end{array}
\end{equation}
where the range $R$ is defined as
\begin{equation}\nonumber
R = \left\{
  \begin{array}{l l}
    (\beta_m,\beta_{m+1}]\text{, if $m<z$}&\\
    &\\
    (\beta_m,F]\cup[1,\beta_1]\text{,  if $m=z$.}
  \end{array} \right.
\end{equation}
\noindent
Additionally, each $S_{ij}(q,\beta_m)$ must satisfy the order enforcement property from the t-frame definition. The s-frame $S_{ij}$ is computed using {\bf Algorithm \ref{alg2}}. The t-frame update algorithm by $j$ upon receiving $S_{ij}$ is shown in {\bf Algorithm \ref{alg3}}.

\begin{algorithm}[t]
\caption{Updating routing table $T_j$ upon receiving $S_{ij}$}
\label{alg3}
\begin{algorithmic}
\FOR {$q=1\hdots K$}
\FOR {$p=1\hdots F$}
\IF {$T_j(q,p)=\varnothing$ {\bf or} $S_{ij}(q,p)<T_j(q,p).\delta$}
\STATE $T_j(q,p).\delta:=S_{ij}(q,p);$
\STATE $T_j(q,p).\xi:=i;$
\ENDIF
\ENDFOR
\ENDFOR
\STATE Enforce descending order of delays down a column
\end{algorithmic}
\end{algorithm}

Let us illustrate the route dissemination process discussed above through an example. Assume that nodes $A$ and $B$ meet the destination $D$ directly, and $C$ learns about the available 2-hop paths to $D$ through them. Let $F=6$ slots, i.e., each t-frame or s-frame has dimension $3\times 6$. However, since $B$ and $C$ do not meet, only paths of length up to $2$ are available. When $A$ meets $D$ (Fig. ~\ref{fig_reeper_example}(a)), $T_A$ gets initialized at hop $1$ with the tuple $\langle 0,D\rangle$ at each meeting (colored) slot $\beta_m$ in $M_{AD}$. Similarly, when $B$ meets $D$ (Fig. ~\ref{fig_reeper_example}(b)), $T_B$ gets initialized. Without loss of generality, assume $C$ meets $A$ before $B$. As shown in Fig. ~\ref{fig_reeper_example}(c), $C$ creates $S_{AC}$ using {\bf Algorithm \ref{alg2}} that advertises the best $2$-hop path at each meeting slot $\beta_m$ till the next meeting in $M_{AC}$. Upon receiving this information, $C$ initializes $T_C$. When C meets B, $T_C$ gets updated with any new or 
better information obtained from $S_{BC}$ using {\bf Algorithm \ref{alg3}}. For example, the third slot (highlighted) in Fig. ~\ref{fig_reeper_example}(d) has potentially two conflicting 2-hop paths, and $T_C$ stores $\langle 1,B\rangle$, the least-delay path information.

The forwarding algorithm searches the t-frame in a row-major fashion to find the next-hop node with least cost and best delay that satisfies the remaining deadline for a packet, given the present time $x$ that is relativized to the start of the frame using Equation \ref{eqn1}. Clearly, if a higher-hop path had more delay than a lower-hop path originating in the same slot, the latter would always be selected over the former, since the search proceeds in an increasing order of hops. Thus, storing the former's details would be tantamount to redundant information. Keeping this in mind, the order enforcement has been implemented in {\bf Algorithm \ref{alg2}} and {\bf Algorithm \ref{alg3}}.
\begin{figure}[t]
            \centering
            \begin{subfigure}[b]{0.2\textwidth}
                    \includegraphics[width=\textwidth]{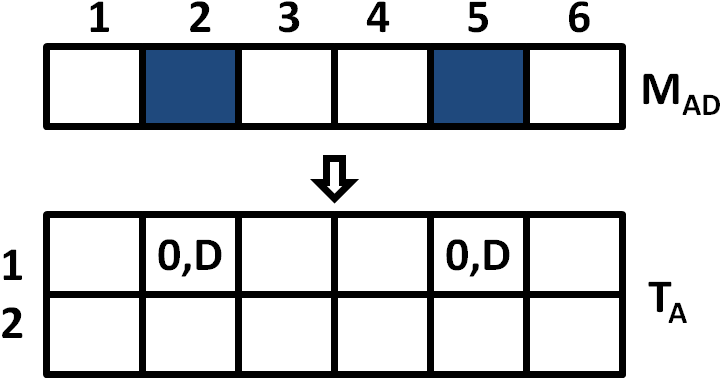}
                    \caption{A meets D,initializes t-frame}
            \end{subfigure}%
            ~ 
            \begin{subfigure}[b]{0.2\textwidth}
                    \includegraphics[width=\textwidth]{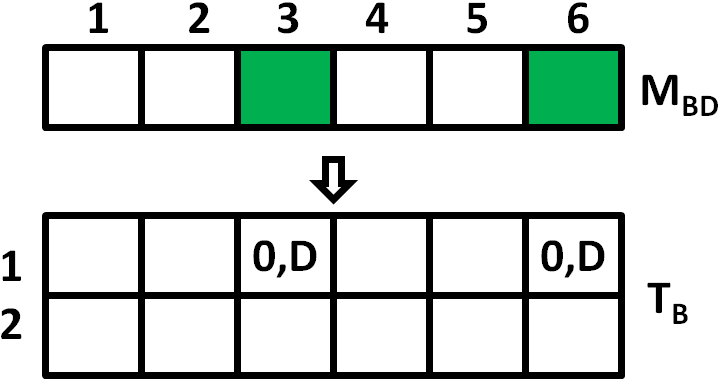}
                    \caption{B meets D,initializes t-frame}
            \end{subfigure}
	    ~
            \begin{subfigure}[b]{0.2\textwidth}
                    \includegraphics[width=\textwidth]{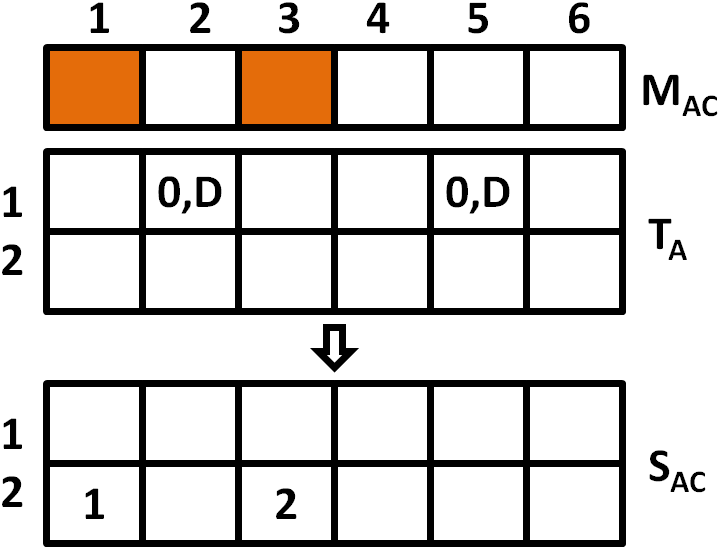}
                    \caption{A meets C,sends s-frame}
            \end{subfigure}
            ~
            \begin{subfigure}[b]{0.2\textwidth}
                    \includegraphics[width=\textwidth]{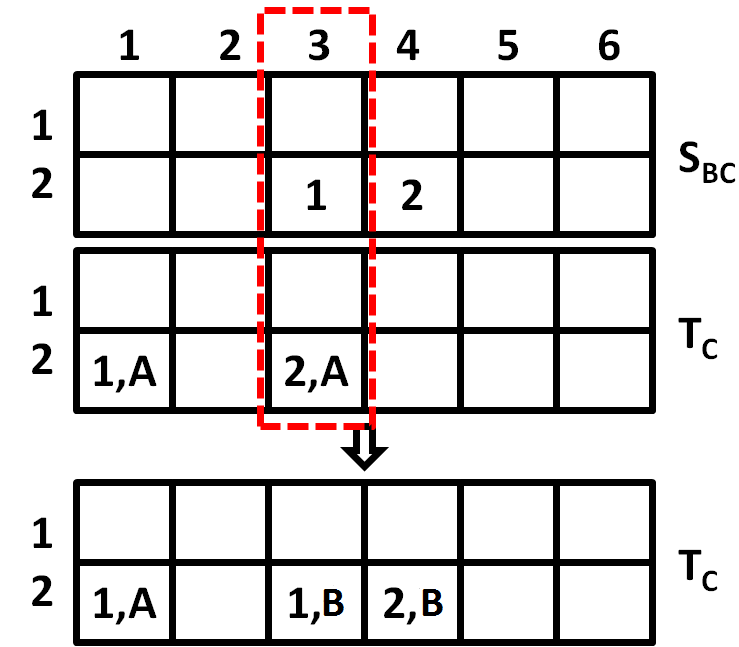}
                    \caption{C meets B,updates t-frame}
            \end{subfigure}
            \caption{An example of path propagation in REAPER}\label{fig_reeper_example}
\end{figure}

{\bf Discussion.}~~~The REAPER framework solves a temporal formulation of the traditional constrained shortest path problem in static graphs, using a multiple metric Bellman Ford algorithm\cite{bellford}. REAPER achieves completeness and optimality (cf. Appendix A) by solving the problem essentially over $F$ successive dynamic graphs in time. This bound, coupled with the small-world property of the underlying network, greatly reduces the complexity in REAPER when compared to competitive protocols such as BUBBLE Rap\cite{Bubble11} (cf. Section \ref{sec_expt}). Since the use of a frame inherently preserves the small world property, a t-frame storing polylogarithmic path lengths is sufficient to achieve near-complete enumerations of best paths. Since $\mathcal{O}(\log{}N)$ space is needed in each t-frame slot for the delay and next hop information, the overall storage complexity per node is $\mathcal{O}((\log{}N)^2)$. The storage is further reduced by order enforcement, and can even be $\mathcal{O}(\log{}N)$ in 
well-behaved networks where delays are an increasing function of path lengths. The control overhead (s-frame) per contact, can correspondingly vary between $\mathcal{O}(\log{}N)$ and $\mathcal{O}(1)$ ($\mathcal{O}(N\log{}N)$ and $\mathcal{O}(N)$ for any-to-any communication). Note that the routing framework does not assume a pre-defined fixed deadline, therefore every packet can have a different deadline and the same t-frame would be used by the forwarding algorithm to serve them all. Importantly, minor variations on the same t-frame can be used for solving a variety of optimization objectives, such as smallest delay--shortest length, highest reliability--shortest length, etc. in a fine-grained way.
\documentclass[main.tex]{subfiles}
\section{Evaluation}
\label{sec_expt}

We have validated our protocol using ns-2.34 simulations over synthetic mobility traces generated for a synthetic \emph{University Campus Environment} mobility scenario, that exhibits the small world phenomenon. This is a $30$-node synthetic trace that uses the Time Variant Community Model (TVCM) of Wei-Jen Hsu et. al. \cite{TVCM09}. For details on the simulation area and TVCM instantiation parameters, please refer to Appendix \ref{App:AppendixB}.

\subsection{Reference Routing Algorithms}
The case of replication-based protocols is relatively straightforward:  as the network is driven towards near-capacity, the performance of many multiple-copy protocols\cite{Epidemic00,SprayWait05,MaxProp06} breaks down well before reaching the regime of ultra-heavy load. We do not therefore evaluate these protocols more comprehensively here. With respect to the single-copy protocols, we eschew consideration of those popular link-metric based single-copy protocols that use some notion of global knowledge of contacts or infrastructure such as well-known landmarks \cite{Per09,OptimalProbForwarding09}. Instead, we focus on the two canonical representatives of infrastructure-free protocols: MEED-DVR, an optimization algorithm based on the ``Minimum Estimated Expected Delay (MEED)'' \cite{MeedDvr07} metric; and ``Probabilistic Routing Protocol using a History of Encounters and Transitivity (PROPHET)'' \cite{Prophet04}, which uses incremental forwarding based on node meeting probabilities that are aged according to 
three parameters (the designers' parameter choices have been honored in this evaluation). For sociability metric-based protocols,  we focus on the canonical state-of-the-art representative, BUBBLE Rap\cite{Bubble11}, which  uses node communities and moving average degrees to route to the destination.

We instantiate four versions of REAPER, with packet deadlines of $96$, $72$, $48$ and $24$ hours, to evaluate the sensitivity to deadline selection in the simulations. The simulation spans $7$ virtual days. Between the second and third days, the network generates $500$-byte packets at various rates between $1$ bps to $9.6$ Kb/s for $86300$ seconds. The network has the ns-2 default link capacity of $1$ Mbps, even though the network is connected to the base station only about $1\%$ of the time, and the actual achievable capacity is estimated to be lower than $10$ Kbps. Thus, the traffic rates range from ultra-low to ultra-high for this network.

\subsection{Results}
\textbf{Average Path Length (Cost).}~~All four versions of REAPER have substantially lower average path lengths (Fig.~\ref{fig_TVCM}(a)). As expected, PROPHET has the highest number of transmissions (about $22\times$ that of REAPER at the heaviest load), since the other two protocols globally optimize their respective metrics. Further, REAPER exploits the small world to yield logarithmic path lengths and is about $4.5\times$ better than MEED-DVR.
\begin{figure}[t]
            \centering
            \begin{subfigure}[b]{0.35\textwidth}
                    \includegraphics[width=\textwidth]{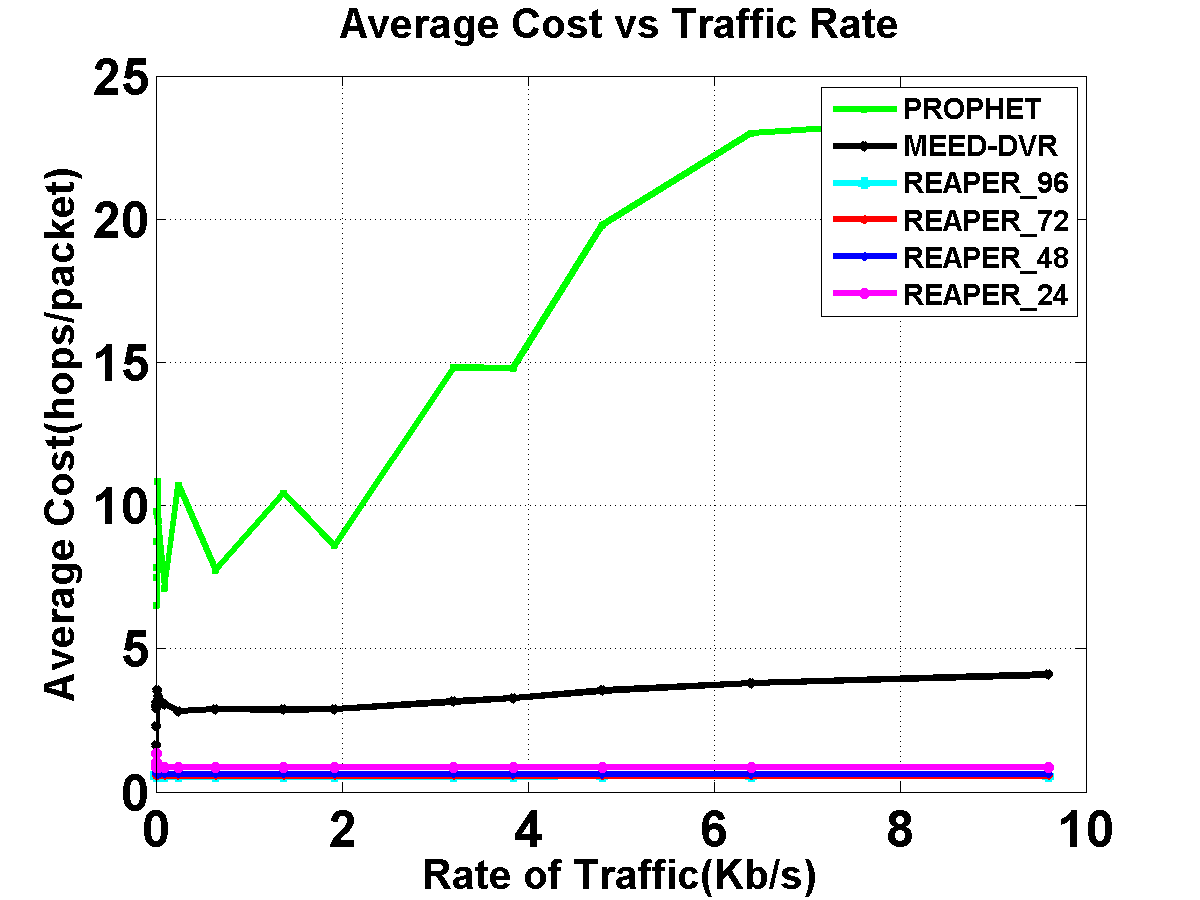}
                    \caption{Average cost}
            \end{subfigure}%
            ~ 
            \begin{subfigure}[b]{0.35\textwidth}
                    \includegraphics[width=\textwidth]{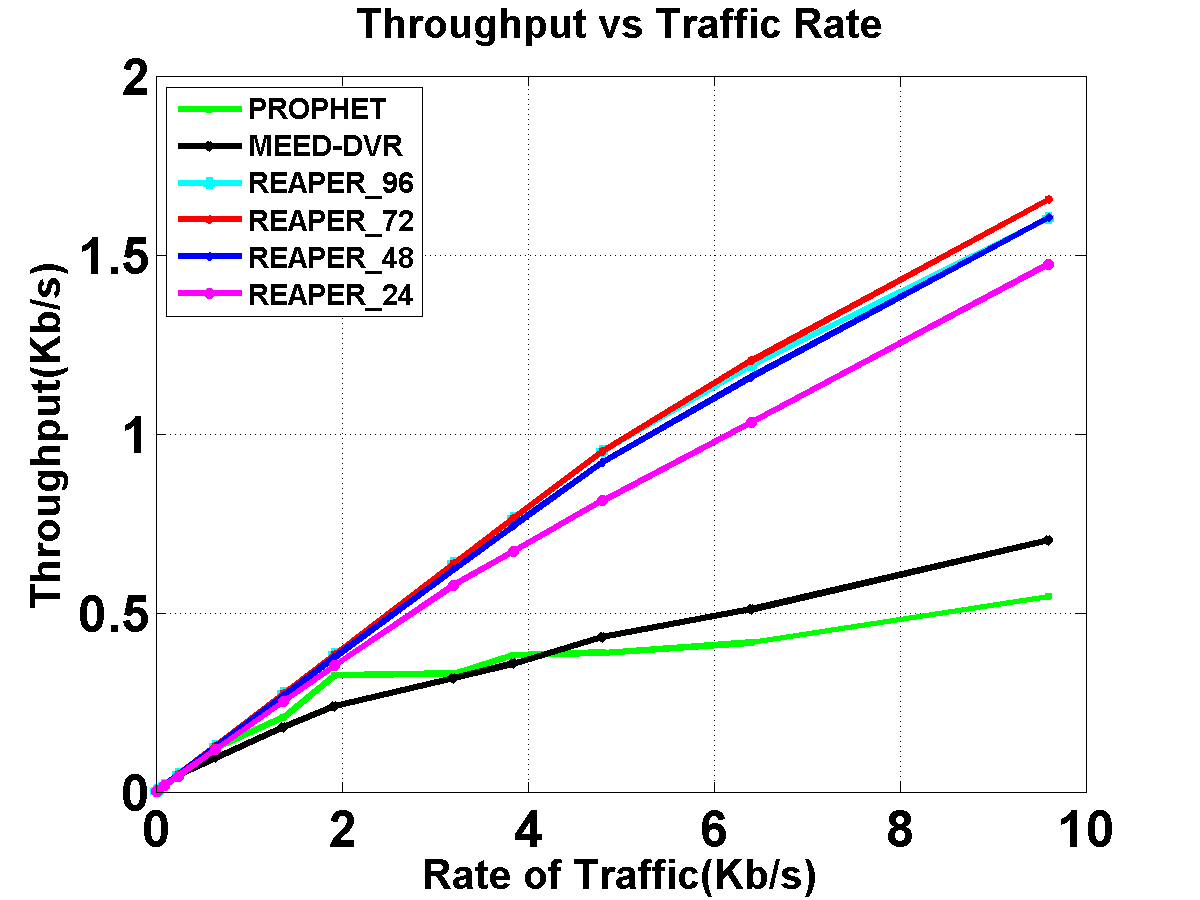}
                    \caption{Throughput}
            \end{subfigure}
	    
            \begin{subfigure}[b]{0.35\textwidth}
                    \includegraphics[width=\textwidth]{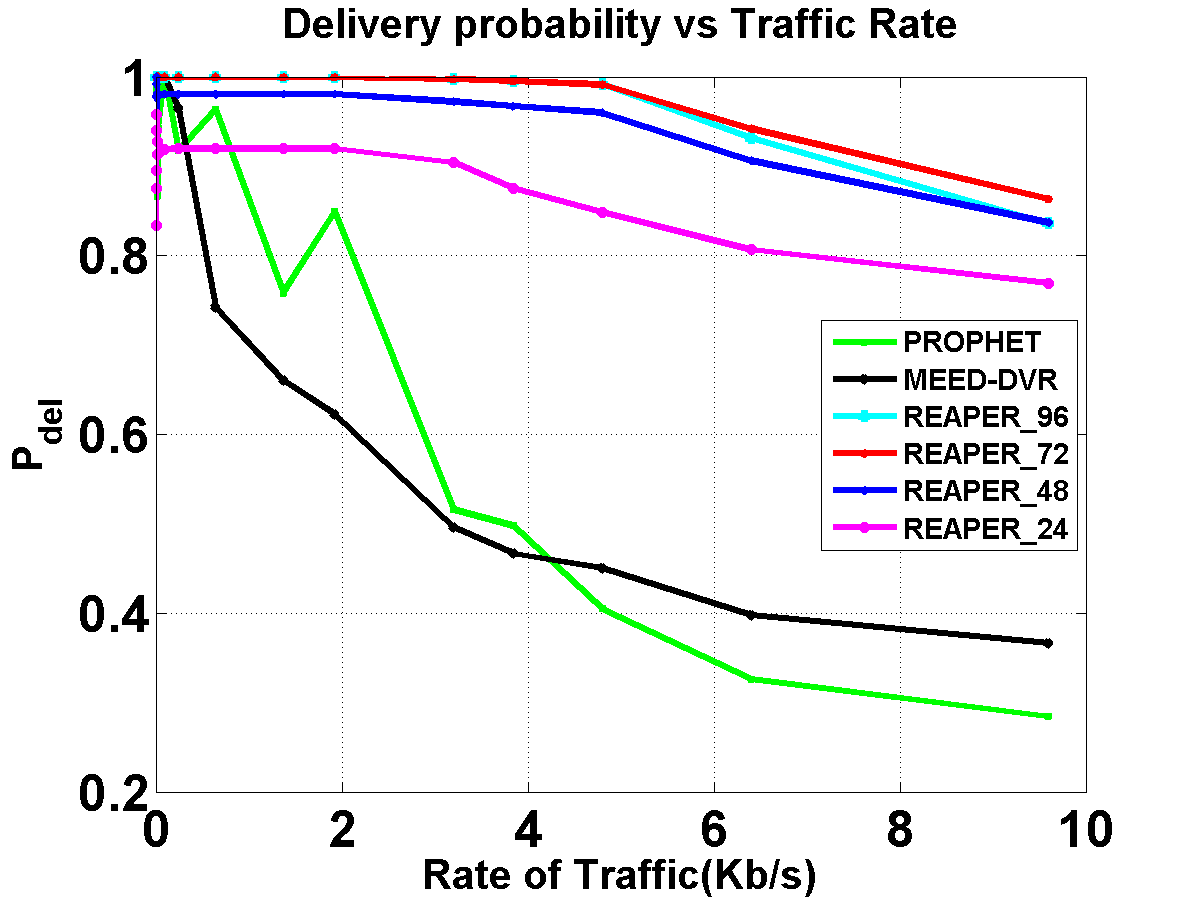}
                    \caption{Delivery probability}
            \end{subfigure}
            ~
            \begin{subfigure}[b]{0.35\textwidth}
                    \includegraphics[width=\textwidth]{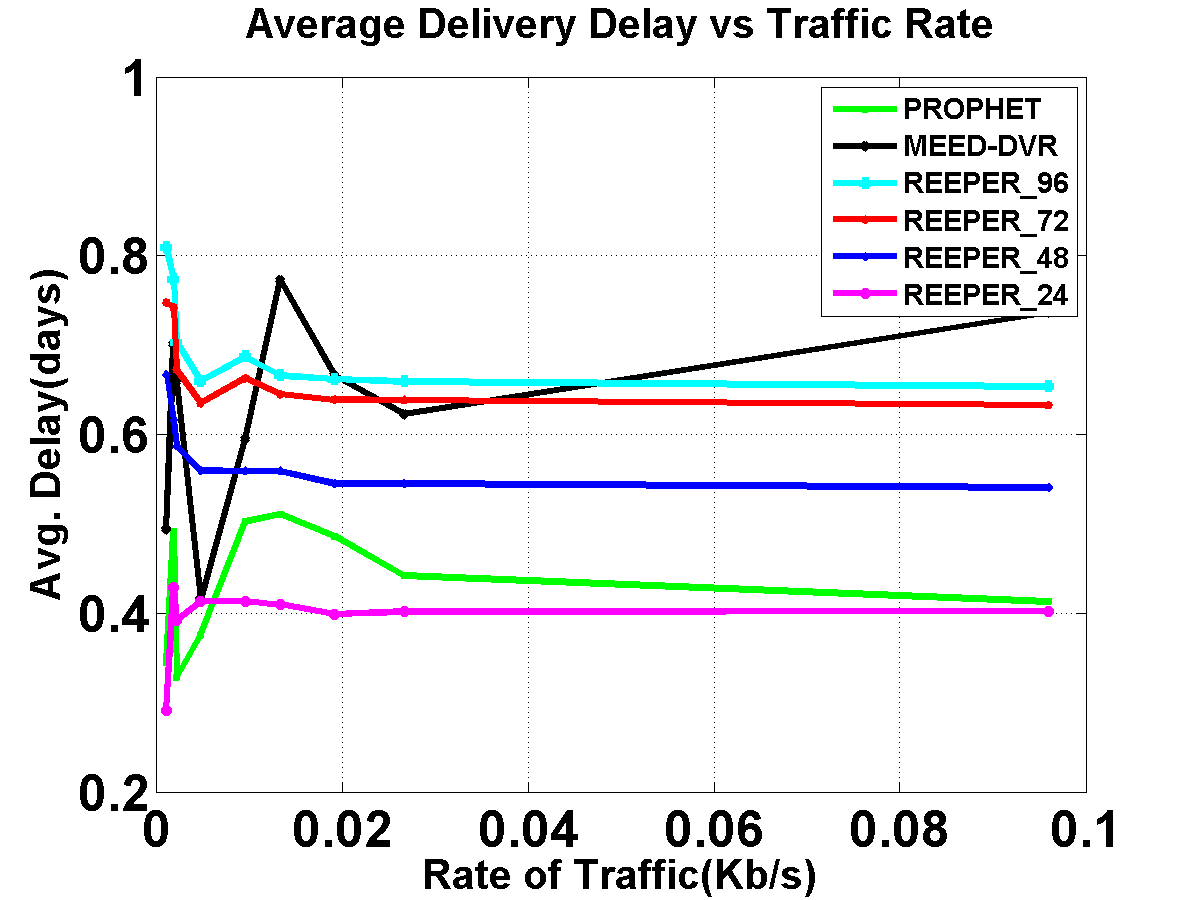}
                    \caption{Average delivery delay (low-traffic regime)}
            \end{subfigure}
            \caption{Experimental results in the campus environment}\label{fig_TVCM}
\end{figure}

\textbf{Throughput and Delivery Probability.}~~As observed in Fig.~\ref{fig_TVCM}(b), at ultra-high load (i.e. {\bf 9.6} Kb/s) REAPER\_96, REAPER\_72, REAPER\_48 and REAPER\_24 respectively have {\bf 135\%, 128\%, 128\%, 114\%} higher throughput than MEED-DVR and {\bf 200\%, 175\%, 175\%, 150\%} higher throughput than PROPHET. Observe that PROPHET and MEED-DVR lose capacity much quicker than all four versions of REAPER. A consequence of this is the loss in their delivery probabilities (Fig.~\ref{fig_TVCM}(c)). In general, the four flavors of REAPER perform increasingly better with increasing deadlines. This is because shorter deadlines entail the use of longer paths to the destination and also result in higher packet drops due to possibly early expiration of deadlines.

\textbf{Average Delivery Delay.}~~Fig.~\ref{fig_TVCM}(d) plots the average delivery delays in the low traffic regime (up to {\bf 0.096} Kb/s) where all protocols have a {\bf 100\%} delivery ratio\footnote{We limit comparison to the low-traffic regime to get a fair comparison: the average delivery delay metric gets biased once packet losses start occurring.}. We see that REAPER\_96 and REAPER\_72 perform comparably with MEED-DVR, while REAPER\_48 and REAPER\_24 have lesser average delays in the expected order of their deadlines. The good delay performance in REAPER despite deadlines is once again attributable to the use of time-dependent paths.

\textbf{Comparison with Sociability-Based Routing}~~We now evaluate REAPER in comparison with the social routing benchmark Distributed BUBBLE Rap (DiBUBB)\cite{Bubble11}, a protocol that classifies nodes into distinct communities and uses the average node degree over the past $6$ hours as a measure of individual centralities. Routing proceeds in an incremental fashion till the destination class is reached, and similarly to the destination node by forwarding to higher-degree members in the same class or to the base station itself. Because community detection is computationally expensive and incurs considerable overhead, we use an optimized version of the proposed algorithm SIMPLE\cite{Bubble11} for our evaluation. 

Like DiBUBB, REAPER exploits the small-world to achieve high throughput using logarithmic average path costs, albeit with traditional link metrics. In fact, REAPER uses slightly higher average path costs to the destination ($1.86$ vs $1.3$ with packet deadlines of $24$ hours, and $1.59$ vs $1.24$ with deadlines of $48$ hours in the regime of ultra-heavy traffic). This slightly increased costs, however, results in noticeable benefits: REAPER achieves respectively about {\bf 42\%}, {\bf 30\%} and {\bf 10\%} higher throughput with deadlines of $12$, $24$ and $48$ hours respectively. This result hints strongly at the benefits of fine-grained path planning: REAPER can change the planned path on the fly based on the remaining time for a packet, and maintain throughput by potentially forwarding packets that are near expiration over slightly longer routes. Social-based forwarding is oblivious to this dynamic change in urgency of a packet based on its remaining time, and uses the exact same paths for all packets 
regardless of their priorities (deadlines). In terms of overhead, much greater benefits are observed. Since DiBUBB always needs to transfer $\mathcal{O}(n\log{}n)$ community and familiarity sets per contact, REAPER has an $\mathcal{O}(n)$ to $\mathcal{O}(n\log{}n)$ times lower overhead than the former.
\documentclass[main.tex]{subfiles}
\section{Conclusion}
\label{sec_conclusion}
In this work, we have demonstrated the benefits of time-dependent path planning in the context of DTN routiung. We have designed a low-complexity online mechanism for predicting the pairwise meeting schedules of nodes, without explicitly estimating the underlying mobility statistics. By exploiting the ability to predict, we have designed a low overhead, self-stabilizing routing framework that performs fine-grained optimization of path hops under deadline constraints per se, but can be used for a variety of other optimization metrics. The delayed delivery mechanism coupled with time-sensitive global optimization of routing costs makes REAPER tunable to a wide range of capacity-delay trade-offs. As future work, we plan to extend the framework to encompass socially derived metrics to achieve scalability in large-scale human DTNs.

   \bibliography{dtnbib}

\begin{thebibliography}{10}

\bibitem{PowerLaw07}
A.~Chaintreau, P.~Hui, et~al.
\newblock Impact of human mobility on opportunistic forwarding algorithms.
\newblock {\em IEEE Trans. on Mobile Computing}, 6(6), 2007.

\bibitem{Dichotomy07}
T.~Karagiannis, J.-Y.~Le Boudec, et~al.
\newblock Power law and exponential decay of inter contact times between mobile
  devices.
\newblock In {\em ACM MobiCom}, 2007.

\bibitem{Pairwise07}
V.~Conan, J.~Leguay, et~al.
\newblock Characterizing pairwise inter-contact patterns in delay tolerant
  networks.
\newblock In {\em ICST Autonomics}, 2007.

\bibitem{Watts98}
D.J. Watts and S.H. Strogatz.
\newblock Collective dynamics of small-world networks.
\newblock {\em Nature}, 393(6684), 1998.

\bibitem{Kleinberg00}
J.~Kleinberg.
\newblock Navigation in a small world.
\newblock {\em Nature}, 406(6798), 2000.

\bibitem{Nguyen12}
A.-D. Nguyeng, P.~Senac, et~al.
\newblock Understanding and modeling the small-world phenomenon in dynamic
  networks.
\newblock In {\em ACM MSWiM}, 2012.

\bibitem{MeedDvr07}
E.P.C. Jones, L.~Li, et~al.
\newblock Practical routing in delay-tolerant networks.
\newblock {\em IEEE Transactions on Mobile Computing}, 6(8), 2007.

\bibitem{Prophet04}
A.~Lindgren, A.~Doria, et~al.
\newblock Probabilistic routing in intermittently connected networks.
\newblock {\em ACM SIGMOBILE Mobile Computing and Communications Review}, 7(3),
  2003.

\bibitem{Bubble11}
P.~Hui, J.~Crowcroft, et~al.
\newblock {BUBBLE} {R}ap: {S}ocial-based forwarding in delay-tolerant networks.
\newblock {\em IEEE Trans. on Mobile Computing}, 10(11), 2011.

\bibitem{Clauset07}
A.~Clauset and N.~Eagle.
\newblock Persistence and periodicity in a dynamic proximity network.
\newblock In {\em DIMACS/DyDAn Workshop on Computational Methods for Dynamic
  Interaction Networks}, 2007.

\bibitem{Nguyen13}
A.-D. Nguyeng, P.~Senac, et~al.
\newblock How disorder impacts routing in human-centric disruption tolerant
  networks.
\newblock In {\em ACM SIGCOMM Workshop on Future Human-centric Multimedia
  Networking}, 2013.

\bibitem{Epidemic00}
A.~Vahdat and D.~Becker.
\newblock Epidemic routing for partially-connected ad hoc networks.
\newblock Technical report, 2000.

\bibitem{SprayWait05}
T.~Spyropoulos, K.~Psounis, et~al.
\newblock Spray and wait: {An} efficient routing scheme for intermittently
  connected mobile networks.
\newblock In {\em ACM WDTN}, 2005.

\bibitem{MaxProp06}
J.~Burgess, B.~Gallagher, et~al.
\newblock Maxprop: {Routing} for vehicle-based disruption-tolerant networks.
\newblock In {\em IEEE INFOCOM}, 2006.

\bibitem{Rapid07}
A.~Balasubramanian, B.~Levine, et~al.
\newblock {DTN} routing as a resource allocation problem.
\newblock {\em ACM SIGCOMM Computer Communication Review}, 37(4), 2007.

\bibitem{BWAR12}
M.~Alresaini, M.~Sathiamoorthy, et~al.
\newblock Backpressure with adaptive redundancy ({BWAR}).
\newblock In {\em IEEE INFOCOM}, 2012.

\bibitem{Per09}
Q.~Yuan, I.~Cardei, et~al.
\newblock Predict and relay: {An} efficient routing in disruption-tolerant
  networks.
\newblock In {\em ACM MobiHoc}, 2009.

\bibitem{OptimalProbForwarding09}
C.~Liu and J.~Wu.
\newblock An optimal probabilistic forwarding protocol in delay tolerant
  networks.
\newblock In {\em ACM MobiHoc}, 2009.

\bibitem{Simbet}
E.~M. Daly and M.~Haahr.
\newblock Social network analysis for routing in disconnected delay-tolerant
  {MANET}s.
\newblock In {\em ACM MobiHoc}, 2007.

\bibitem{Gao10}
W.~Gao and G.~Cao.
\newblock On exploiting transient contact patterns for data forwarding in delay
  tolerant networks.
\newblock In {\em IEEE ICNP}, 2010.

\bibitem{Zhang13}
X.~Zhang and G.~Cao.
\newblock {Transient community detection and its application to data forwarding
  in delay tolerant networks}.
\newblock In {\em IEEE ICNP}, 2013.

\bibitem{haggle06}
J.~Scott, R.~Gass, et~al.
\newblock {CRAWDAD} data set cambridge/haggle (v. 2006-01-31).
\newblock http://crawdad.org/cambridge/haggle/, 2006.

\bibitem{pmtr08}
P.~Meroni and S.~Gaito andothers.
\newblock {CRAWDAD} data set unimi/pmtr (v. 2008-12-01).
\newblock http://crawdad.org/unimi/pmtr/, 2008.

\bibitem{bellford}
E.~Tardos and J.~Kleinberg.
\newblock Algorithm design.
\newblock Pearson/Addison-Wesley, 2008.

\bibitem{TVCM09}
W.-J. Hsu, T.~Spyropoulos, et~al.
\newblock Modeling spatial and temporal dependencies of user mobility in
  wireless mobile networks.
\newblock {\em IEEE/ACM Transactions on Networking}, 17(5), 2009.

\end{thebibliography}
\documentclass[main.tex]{subfiles}
\appendix
\section{APPENDIX: Analysis of REAPER} \label{App:AppendixA}
\subsection{Complete Routing Module}
In this section, we formally present the complete routing module using the standard guarded-command notation of distributed programs (\textbf{Algorithm \ref{prog}}).  We let $\mathbb{V}$ represent the set of relay nodes. Also, for the sake of generality, the destination node $D$ is assumed to have a t-frame with only a $0-$hop path to itself at all times. Each relay $i$ additionally maintains $S^{o}_{ji}$, a single history of the last s-frame sent to it by every neighbor $j$. The program actions $(1)$-$(6)$ are executed assuming a weakly fair scheduling policy. Actions $(4)$ and $(6)$ respectively create an s-frame to be sent to a neighbor, and incorporate a t-frame received from a neighbor using algorithms previously discussed in Section IV. Actions $(1)$-$(3)$ and $(5)$ are required to ensure that the program self-stabilizes after any errors, as explained next.

\begin{algorithm}[tp]
\caption{REAPER Routing Protocol \emph{Process $i$}}
\label{prog}
\begin{algorithmic}
\STATE \textbf{\emph{Guard macro definitions:}}
\STATE $(A)\quad C.1\equiv[(i=D)\implies((T_i(0,1).\langle \delta,\xi\rangle=\langle0,D\rangle)\land$
\STATE \qquad$(\forall q\in [1,K],p\in [1,F]: T_i(q,p).\langle \delta,\xi\rangle=\langle \infty,X\rangle))]\land$
\STATE \qquad$[(i\neq D)\implies((\forall q,p: q\in[1,K], p\in[1,F]:$
\STATE \qquad$ T_i(q,p)\delta>0\land T_i(q,p)\xi\in [(\mathbb{V}-\{i\})\cup \{D\}])\land$
\STATE \qquad\qquad$(\forall p: p\in [1,F]: T_i(0,p).\langle \delta,\xi\rangle=\langle \infty,X\rangle))]$
\newline

\STATE $(B)\quad C.2\equiv (i\neq D)\implies(\forall q\in [1,K],p\in [1,F]:$
\STATE \qquad$(T_i(q,p).\delta=\infty\equiv T_i(q,p).\xi=X)\land$
\STATE \qquad$(T_i(q,p).\delta<\infty\implies(T_i(q,p).\delta=S^{o}_{(T_i(q,p).\xi)i}.\delta)\land$
\STATE \qquad\qquad$(M_{i(T_i(q,p).\xi)}(p)=1)\land((q>1)\implies$
\STATE \qquad\qquad\qquad$(T_i(q,p)<T_i(q-1,p)))$
\newline

\STATE $(C)\quad C.3\equiv$ Received $S_{ji}$ from $j\implies$
\STATE \qquad $\lnot (\exists q\in[1,K], p\in[1,F]: (T_i(q,p).\xi=j)\land$
\STATE \qquad \qquad $(T_i(q,p).\delta<S_{ji}(q,p).\delta)))$
\newline

\STATE \textbf{\emph{Program actions:}}
\STATE $(1)\quad(i=D)\land\lnot C.1$\quad $\longrightarrow$ \quad $T_i(0,1).\langle \delta,\xi\rangle:=\langle 0,D\rangle;$
\STATE \qquad $(||q,p: q\in[1,K], p\in[1,F]:$ Reset $T_i(q,p)$
\newline

\STATE $(2)\quad(i\neq D)\land\lnot C.1$\quad $\longrightarrow$\quad Reset $T_i$
\newline

\STATE $(3)\quad(i\neq D)\land C.1 \land\lnot C.2$\quad $\longrightarrow$
\STATE \qquad $(||m\in [1,F], n\in [1,K]: T_i(n,m).\xi=T_i(q,p).\xi: $ Reset $T_i(n,m)$
\newline

\STATE $(4)\quad(M_{ij}\neq \varnothing)\land(j\neq D)\land C.1\land C.2$\quad $\longrightarrow$
\STATE \qquad Execute \textbf{Algorithm 1}; Send $S_{ij}$ to $j$
\newline

\STATE $(5)\quad(M_{ij}\neq \varnothing)\land(i\neq D) \land$ Receive $S_{ji}$ from $j$ $\land C.1 \land C.2 \land \lnot C.3$ \quad $\longrightarrow$
\STATE $(||m\in [1,F], n\in [1,K]: T_i(n,m).\xi=j: $ Reset $T_i(n,m)$
\newline

\STATE $(6)\quad(M_{ij}\neq \varnothing)\land(i\neq D)\land$ Receive $S_{ji}$ from $j$ $\land C.2 \land C.3$ \quad $\longrightarrow$\quad Execute \textbf{Algorithm 2}; $S^{o}_{ji} := S_{ji};$
\end{algorithmic}
\end{algorithm}

The guard macros $C.1$, $C.2$ are sanity checks for t-frames and s-frames, required for self-stabilization. $C.1$ states that a destination node can only have a $0-$hop path to itself, while all the other nodes can only have $1\hdots K$ hops. Whenever $C.1$ is not satisfied at a node, the corresponding t-frame is reset by actions $(1)$ and $(2)$. $C.2$ states that an infinite-delay (non-data) slot in a t-frame should have no next hop, and a finite-delay slot should tally with the last-sent s-frame from the corresponding next hop. $C.2$ also checks the descending order of delays at a certain slot in the t-frame, as enforced by \textbf{Algorithm 2}. the If any data slot violates this condition, it would be reset by the action $(3)$.

The condition $C.3$ checks if an s-frame received from a neighbor is legal. The condition states that delay at any slot offered by the new s-frame cannot be higher than what it might have provided earlier for the same hop. This is understandable, since paths are upgraded only if a lower delay is discovered for the same hop in the same slot. When $C.3$ is violated, a node performs what we call the \emph{best-effort quarantine action} (action $(5)$). It eliminates from the t-frame, all contributions from the dubious neighbor in a best effort at limiting the propagation of spurious paths to its other neighbors. However, it may so happen that the spurious s-frame gets forwarded a second time to the same node. Since a single history state is kept, the $C.3$ test would now pass and the spurious frame would get saved. But due to progress properties of the protocol discussed in the next section, eventually a valid s-frame would be forwarded to the node from the same neighbor. It is easy to see that the quarantine 
action would fire at most once during the process of replacement of an invalid s-frame with a new valid s-frame, and from the next meeting onwards $C.3$ would always be satisfied.

\subsection{Correctness and Optimality}
We establish the correctness and optimality of REAPER via the following two lemmas:\\

\noindent 
{\bf Lemma 1.} {\em The information in $T_i$ is both necessary and sufficient for REAPER to find the best cost path for $i$ at any instant $x$.}

\noindent 
{\em Proof Sketch}: We consider the two obligations separately.

\emph{Necessity of $T_i$}: A valid path for a node can be an invalid path for its higher-hop neighbor in terms of deadline satisfiability, in which case the best path could be a different path with a possibly higher cost. Therefore, storing a proper subset of every-hop best paths in $T_i$ can require additional requests for paths to the neighboring nodes since the deadline may not be met through any of them, or can yield suboptimal paths to the destination in terms of cost.

\emph{Sufficiency of $T_i$}: It suffices to show that when the routing algorithm for REAPER converges, the best path $\widehat{L}_i(t)$ at any instant can be computed from $T_i$ by the forwarding algorithm. Note that $\widehat{L}_i(t)$ must have a valid suffix\\$\widehat{L}_{\xi(\widehat{L}_i(t),i)}[\psi(\widehat{L}_i(t),i)]$ from its next hop relay $\xi(\widehat{L}_i(t),i)$. This information is propagated to $i$ through the s-frame. By way of contradiction, assume that when $i$ meets a relay $\xi(\widehat{L}_i(t),i)$ in slot $\beta_{m_1}$, the best $l$-hop path information through $\xi(\widehat{L}_i(t),i)$ till the next meeting (at slot $\beta_{m_1+1}$) is not forwarded by $r$. Since $S_{\xi(\widehat{L}_i(t),i)i}$ compares every $(l-1)$- hop path through $r$ in the interval $(\beta_{m_1},\beta_{m_1+1}]$ to compute the best $l$-hop path at $\beta_{m_1}$, this is not possible. Since the information of all best paths is present in the system, any instant $x$ can be mapped to a specific slot in $T_i$ and the 
corresponding best path computed using the forwarding algorithm.
\hfill{$\Box$}\\

\noindent 
{\bf Lemma 2.} {\em REAPER's routing and forwarding algorithms satisfy Equation \eqref{eqopt}.}

\noindent 
{\em Proof Sketch}:
Assume that a path $L_i(x)$ is chosen at time $x$ by node $i$. Recall its cost is $\theta(L_i(x))$ hops, its next hop is $\xi(L_i(x),i)=r'$. Let the contact with $r'$ happen at meeting slot $\beta_{m^\prime}$. By Lemma 1, the routing algorithm ensures the optimal path $\widehat{L}_i(x)$ is in $T_i$, and since the forwarding algorithm first checks for the lowest cost path, and then for lowest delay option, it cannot be that  \emph{$\theta_{\widehat{L}_i(t)} \neq \theta_{L_i(x)}$}. If $L_i(x)$ is sub-optimal, we have the following cases:

\emph{Case I. The optimal path $\widehat{L}_i(x)$ is a $\theta(L_i(x))$-hop path through the same relay $r'$ at a different meeting slot $\beta_m$}.~~Since $L_i(x)$ is sub-optimal, we must have
$$\delta_{L_i(x)}>\delta_{\widehat{L}_i(x)}$$
But since the forwarding algorithm compares same-hop best-delay paths at all slots in $T_i$ and minimizes $\delta_{L_i(x)}$, and by {\bf Lemma 1}, no path is eschewed, this is impossible.

\emph{Case II. The optimal path $\widehat{L}_i(x)$ is a $\theta(L_i(x))$-hop path through a different relay $r$ at a different meeting slot $\beta_m$}.~~Since $T_i$ stores information corresponding to all relays, the algorithm would choose $\widehat{L}_i(x)$ through $r$ and not $r'$.
\hfill{$\Box$}

\subsection{Self-Stabilization and Progress}
We prove the self-stabilization and progress properties of our distributed routing algorithm with a single base station $D$ for simplicity, using the convergent-stair technique. The extension to a destination class of nodes is straightforward.
\begin {figure}[t]
\centering
\fbox{
\begin{minipage}[b]{0.97\linewidth}
\begin {tabbing}
$H.0\equiv true$\\
$H.1\equiv (\forall i,j:i,j\in \mathbb{V}\cup {D}: C.1\land C.2\land C.3)$\\
$H.2\equiv H.1\land(\forall i,j:i\in I_1,j\in I_0:M_{ij}\neq \varnothing\implies$\\
\qquad$(\forall q=1,p\in [1,F]: T_i(q,p).\delta\leq S_{ji}(q,p).\delta$))\\
\vdots\\
$H.(r+1)\equiv H.r\land(\forall i,j:i\in I_r,j\in I_{r-1}:M_{ij}\neq \varnothing$\\
\qquad$\implies(\forall q\in [1,r],p\in [1,F]: T_i(q,p).\delta\leq S_{ji}(q,p).\delta$))\\
\vdots\\
$H.(K+1)\equiv H.K\land(\forall i,j:i\in I_K,j\in I_{K-1}:M_{ij}\neq \varnothing$\\
\qquad$\implies(\forall q\in [1,K],p\in [1,F]: T_i(q,p).\delta\leq S_{ji}(q,p).\delta$))
\end{tabbing}
\end {minipage}
}
\caption{REAPER Invariants}
\label{fig_inv}
\end {figure}

Let $I_r, r\in[0,K]$ be the set of nodes having an $r$-hop path to the destination. It is easy to see that $I_0=\{D\}$, a singleton set. Further, it is not necessary that $I_l\cap I_m=\varnothing$ for any $l,m\in(0,K]$. We then define the following set of invariants as shown in Fig.~\ref{fig_inv}. It may be observed that $H.(K+1)$ implies stabilization of routes in the system.

\subsubsection{Closure Properties} We first prove the closure properties of the invariants in Fig. \ref{fig_inv}.

\noindent
(A)~\emph{$H.1$ is closed}: When $H.1$ holds for a process $i$, actions $(1)$,$(2)$,$(3)$ and $(5)$ are disabled. We then have the following two cases:

\emph{Case I. $i=D$}: In this case, only action $(4)$ is enabled. Since no change is made to $T_D$ through the action, all the three sub-clauses $C.1, C.2$ and $C.3$ continue to hold and $H.1$ is preserved.

\emph{Case II. $i\neq D$}: Following the above logic, since the send action $(4)$ makes no changes to $T_i$, $H.1$ is preserved. For the receive action $(6)$, even if an existing value in $T_i$ is overwritten, note that $(6)$ fires only when the received frame is well-behaved with respect to $C.3$, and the corresponding send happens when both $C.1$ and $C.2$ hold. Since each of the clauses is conservative, the changed $T_i$ also preserves $H.1$.

\noindent
(B)~\emph{$H.(r+1)$ is closed $\forall r\in[1,K]$}: We prove this by the method of induction.

\noindent
\emph{Base case}: We have already shown that $H.1$ is closed. Therefore, it suffices to show that the second clause is closed given $H.1$. Note that $I_0=\{D\}$ and that hop $0$ is always stable. Given $H.1$, therefore, $\forall i\in I_1$, $S_{Di}$ never changes. For each $i$, the send action $(4)$ at $i$ is disabled for $j=D$. The receive action $(6)$ can only overwrite $T_i(1,.)$ the first time $i$ meets $D$, and the predicate holds by virtue of equality. If $T_i$ is not overwritten by action $(6)$, the predicate is trivially preserved. Thus, $H.2$ is closed.

\noindent
\emph{Induction hypothesis}: $H.r$ is closed.

\noindent
\emph{Induction}: We need to show that $H.(r+1)$ is closed. For either $i$ or $j$, $H.r$ implies only actions $(4)$ and $(6)$ are enabled. The send actions $(4)$ at $i$ or $j$ do not affect $T_i$ or $S_{ji}$ till the $r$\textsuperscript{th} hop, so the predicate is trivially preserved. For process $j$, the receive action $(6)$ does not affect either $T_i$ or $S_{ji}$. For process $i$, since $H.r$ holds, no improved information till the $r$\textsuperscript{th} hop can be obtained from $j$ through the receive action $(6)$, and the predicate continues to hold. Thus $H.(r+1)$ is closed.

\subsubsection{Progress Properties} We now prove that our algorithm self-stabilizes, i.e. given any state $H.0$, it converges to a state $H.(K+1)$ given a bounded number of failures. We prove the progress property using the following sequence of \emph{leads-to} ($\longmapsto$) formulations assuming a weakly fair scheduling policy, in the following two steps:

\noindent
(A)~\emph{$H.0\longmapsto H.1$}: Given a state $s$, let us consider the variant function: $\#(s)=\langle C_1(s), C_2(s), C_3(s), C_5(s)\rangle $, a lexicographic ordering where $C_x(s)$ = the number of nodes in N that satisfy the guard of the $x$\textsuperscript{th} action. Clearly, whenever action 1, 2, 3 or 5 executes, the corresponding $C_x(s)$ decreases by $1$; hence the variant function is monotonically decreasing with a lower limit of $\langle 0,0,0,0\rangle$. When $\#(s)$ reaches its lower bound, the system is in state $H.1$.

\noindent
(B)~\emph{$H.r\longmapsto H.(r+1) \forall r\in[1,K]$}: $H.r$ implies that only actions 4 and 6 are enabled. Consider the variant function $V_{r+1}(s) = \Omega_{r+1}-P_{r+1}(s)$, where $\Omega_{r+1}$ is an oracle variable denoting the total number of $r$-hop paths to be eventually recorded in the system, and $P_{r+1}(s)$ is the total number of $r$-hop paths discovered in state $s$. The function is lower-bounded at $0$ and is non-increasing, since it decreases through an updating receive action involving an s-frame $S.j.i$, $i\in I_r$, $j\in I_{r-1}$, and remains constant otherwise. Due to the system assumption of weak fairness, we can say that the value of $V_{r+1}$ eventually decreases given $H.r$, and the system reaches state $H.(r+1)$ in finite time.

Thus, we prove that the algorithm is self-stabilizing under a bounded number of failures.
\hfill{$\Box$}

\subsection{Cycle Avoidance}
In the steady state, the per-link protocol overhead at a node $i$ is polylogarithmic. Storing the entire path instead of just the next hop would have sufficed to eliminate cycles, but at the cost of higher overhead. Instead, eachnode $i$ excludes from s-frame $S_{ij}$, all the paths that have $j$ as the next hop (as discussed before) as a first step at cycle avoidance. Observe that paths with cycles can still exist in the t-frame; nevertheless, we can assert:\\

\noindent 
{\bf Lemma 3.} {\em At any timeslot $x$, any node $i$ will always select a cycle-free path $\widehat{L}_i(x)$ to the destination.}

\noindent 
{\em Proof Sketch}: We consider two cases:

\emph{Case I. A path $L_i(x)$ has a cycle of length $\pi$ with $i$ as a recurring node}.~~Let $L_i(x)$ be the sequence 
{\raggedright $\langle i, C_1, \hdots, C_{\pi-1}, i, R_1, \hdots, R_{l-1}, D\rangle$}, of which $\widehat{L}_i(x)$ is the suffix $\langle i, R_1, \hdots, R_{l-1}, D\rangle$.  Assuming that there are no other paths in $T_i$, the forwarding algorithm has to select between these two paths. Now $\theta(L_i(x))>\theta(\widehat{L}_i(x))$ and
\begin{equation}\nonumber
\begin{array}{l l}
\delta(L_i(x))>\delta(\widehat{L}_i(x))
\hfill{\text{, if }x\in[\psi(L_i(x),i),\psi(\widehat{L}_i(x),i))}\text{,}
&\\
\delta(L_i(x))=\delta(\widehat{L}_i(x))
\text{, otherwise.}
\end{array}
\end{equation}
With either path, since the forwarding algorithm proceeds in increasing order of hop counts, $\widehat{L}_i(x)$ which has $\pi$ fewer hops would be selected over $L_i(x)$.

\emph{Case II. A path $L_i(x)$ has a cycle of length $\pi$ with a relay $j$, $j \neq i$, as a recurring node}.~~Let $L_i(x)$ be the sequence
{\raggedright $\langle i, R_1, \hdots, j, C_1, \hdots, C_{\pi-1}, j,  \hdots, R_{l-1}, D\rangle$}, and $\widehat{L}_i(t)$ be the corresponding linear path $\langle i, R_1, \hdots, j, \hdots, R_{l-1}, D\rangle$. Both paths would feature on the same start slot in $T_i$ at hop indices $l$ and $l+\pi$ respectively. By reasoning as we did in the previous case,  the forwarding algorithm would select $\widehat{L}_i(x)$.

The case of paths that are a combination of cases \emph{I} and \emph{II} is handled similarly.
\hfill{$\Box$}

\section{APPENDIX: TVCM Parameters Used in Section \ref{sec_expt}} \label{App:AppendixB}
The University Campus Environment is generated using the TVCM model\cite{TVCM09} based on two essential entities: “communities” and “time periods”. We choose $18$ communities
in a geographic area of $1\text{ km}\times 1\text{ km}$ that are classified into
$10$ home, $4$ work, $2$ food and $2$ recreation communities (with
one recreation community doubling as a food community).  Every node has a probability
of going to each community at each time period. Also, every
node has an epoch length and an epoch time which signifies
the stationary time of the nodes at those communities. We have
instantiated parameters for the TVC model to faithfully approximate
mobility patterns on a university campus, to obtain a
synthetic University Campus Trace. A
day has been divided into $5$ time periods, and node mobility
in each period is specified in terms of each node going to
a certain community with a specific probability. Every community is $100\text{ m}\times 100\text{ m}$ in dimension. The node mobilities are generated in such a way that, in
a certain time period $\tau$ , a node is given a specific probability
$P_\tau$ to go to a certain community $C_\tau$, as summarized in Table \ref{tvcm_param}.

\begin{table}[h]
\caption{Summary of Mobility Traces} 
\centering 
\begin{tabular}{c c c c} 
\hline\hline 
Time Period & Occurrence & $C_\tau$ & $P_\tau$\\  
($\tau$) & (24-hr. format) & &\\[0.5ex]
\hline 
1 & $7:00-9:00$, $19:00-21:00$ & Recreation & 0.5\\ 
2 & $9:00-13:00$, $15:00-17:00$ & Work & 0.9\\ 
3 & $13:00-15:00$ & Food & 0.33\\ 
4 & $17:00-19:00$ & Recreation & 0.5\\ 
5 & $ 21:00-7:00$ & Home & 0.9\\ 
[1ex] 
\hline 
\end{tabular}
\label{tvcm_param} 
\end{table}

\end{document}